\documentclass[12pt]{article}
\usepackage{amssymb}
\usepackage{amsmath,epsfig,latexsym,cite,graphicx,a4}

\usepackage{epsf,graphics}
\DeclareMathAlphabet{\mathpzc}{OT1}{pzc}{m}{it}
\begin{document}

\def\thefootnote{\fnsymbol{footnote}}
\def\bea{\begin{eqnarray*}}
\def\eea{\end{eqnarray*}}
\def\stroke{\vrule height8pt width0.4pt depth-0.1pt}
\def\topfleck{\vrule height8pt width0.5pt depth-5.9pt}
\def\botfleck{\vrule height2pt width0.5pt depth0.1pt}
\def\Zmath{\vcenter{\hbox{\numbers\rlap{\rlap{Z}\kern
0.8pt\topfleck}\kern 2.2pt
                   \rlap Z\kern 6pt\botfleck\kern 1pt}}}
\def\Qmath{\vcenter{\hbox{\upright\rlap{\rlap{Q}\kern
                   3.8pt\stroke}\phantom{Q}}}}
\def\Nmath{\vcenter{\hbox{\upright\rlap{I}\kern 1.7pt N}}}
\def\Cmath{\vcenter{\hbox{\upright\rlap{\rlap{C}\kern
                   3.8pt\stroke}\phantom{C}}}}
\def\Rmath{\vcenter{\hbox{\upright\rlap{I}\kern 1.7pt R}}}
\def\Z{\ifmmode\Zmath\else$\Zmath$\fi}
\def\Q{\ifmmode\Qmath\else$\Qmath$\fi}
\def\N{\ifmmode\Nmath\else$\Nmath$\fi}
\def\C{\ifmmode\Cmath\else$\Cmath$\fi}
\def\R{\ifmmode\Rmath\else$\Rmath$\fi}
\def\H{{\cal H}}
\def\NN{{\cal N}}
\def\tv{{\tilde{v}}}
\def\vep{{\tilde{\epsilon}}}
\def\te{{\tilde{\vep}}}
\def\sh{{\rm sh}}
\def\cth{{\rm cth}}
\def\th{{\rm th}}
\def\bk{{\bf k}}
\def\br{{\bf r}}

%
%
%
%
\def\oti{{\otimes}}
\def\lb{ \left[ }
\def\rb{ \right]  }
\def\tilde{\widetilde}
\def\bar{\overline}
\def\hat{\widehat}
\def\*{\star}
\def\[{\left[}
\def\]{\right]}
\def\({\left(}      \def\BL{\Bigr(}
\def\){\right)}     \def\BR{\Bigr)}
    \def\BBL{\lb}
    \def\BBR{\rb}
%
%
\def\zb{{\bar{z} }}
\def\zbar{{\bar{z} }}
\def\frac#1#2{{#1 \over #2}}
\def\inv#1{{1 \over #1}}
\def\half{{1 \over 2}}
\def\d{\partial}
\def\der#1{{\partial \over \partial #1}}
\def\dd#1#2{{\partial #1 \over \partial #2}}
\def\vev#1{\langle #1 \rangle}
\def\ket#1{ | #1 \rangle}
\def\rvac{\hbox{$\vert 0\rangle$}}
\def\lvac{\hbox{$\langle 0 \vert $}}
\def\2pi{\hbox{$2\pi i$}}
\def\e#1{{\rm e}^{^{\textstyle #1}}}
\def\grad#1{\,\nabla\!_{{#1}}\,}
\def\dsl{\raise.15ex\hbox{/}\kern-.57em\partial}
\def\Dsl{\,\raise.15ex\hbox{/}\mkern-.13.5mu D}
%
%
\def\ga{\gamma}     \def\Ga{\Gamma}
\def\be{\beta}
\def\al{\alpha}
\def\ep{\epsilon}
\def\vep{\varepsilon}
\def\dep{d}
\def\arc{{\rm Arctan}}
\def\la{\lambda}    \def\La{\Lambda}
\def\de{\delta}     \def\De{\Delta}
 \def\hD{{\Delta}}
\def\om{\omega}     \def\Om{\Omega}
\def\sig{\sigma}    \def\Sig{\Sigma}
\def\vphi{\varphi}
\def\sign{{\rm sign}}
\def\he{\hat{e}}
\def\hf{\hat{f}}
\def\hg{\hat{g}}
\def\ha{\hat{a}}
\def\hb{\hat{b}}
%
%
\def\CA{{\cal A}}   \def\CB{{\cal B}}   \def\CC{{\cal C}}
\def\CD{{\cal D}}   \def\CE{{\cal E}}   \def\CF{{\cal F}}
\def\CG{{\cal G}}   \def\CH{{\cal H}}   \def\CI{{\cal J}}
\def\CJ{{\cal J}}   \def\CK{{\cal K}}   \def\CL{{\cal L}}
\def\CM{{\cal M}}   \def\CN{{\cal N}}   \def\CO{{\cal O}}
\def\CP{{\cal P}}   \def\CQ{{\cal Q}}   \def\CR{{\cal R}}
\def\CS{{\cal S}}   \def\CT{{\cal T}}   \def\CU{{\cal U}}
\def\CV{{\cal V}}   \def\CW{{\cal W}}   \def\CX{{\cal X}}
\def\CY{{\cal Y}}   \def\CZ{{\cal Z}}

\def\Hp{{\mathbb{H}^2_+}}
\def\Hm{{\mathbb{H}^2_-}}

\def\rvac{\hbox{$\vert 0\rangle$}}
\def\lvac{\hbox{$\langle 0 \vert $}}
\def\comm#1#2{ \BBL\ #1\ ,\ #2 \BBR }
\def\2pi{\hbox{$2\pi i$}}
\def\e#1{{\rm e}^{^{\textstyle #1}}}
\def\grad#1{\,\nabla\!_{{#1}}\,}
\def\dsl{\raise.15ex\hbox{/}\kern-.57em\partial}
\def\Dsl{\,\raise.15ex\hbox{/}\mkern-.13.5mu D}
%
%
%
\font\numbers=cmss12
\font\upright=cmu10 scaled\magstep1
\def\stroke{\vrule height8pt width0.4pt depth-0.1pt}
\def\topfleck{\vrule height8pt width0.5pt depth-5.9pt}
\def\botfleck{\vrule height2pt width0.5pt depth0.1pt}
\def\Zmath{\vcenter{\hbox{\numbers\rlap{\rlap{Z}\kern
0.8pt\topfleck}\kern 2.2pt
                   \rlap Z\kern 6pt\botfleck\kern 1pt}}}
\def\Qmath{\vcenter{\hbox{\upright\rlap{\rlap{Q}\kern
                   3.8pt\stroke}\phantom{Q}}}}
\def\Nmath{\vcenter{\hbox{\upright\rlap{I}\kern 1.7pt N}}}
\def\Cmath{\vcenter{\hbox{\upright\rlap{\rlap{C}\kern
                   3.8pt\stroke}\phantom{C}}}}
\def\Rmath{\vcenter{\hbox{\upright\rlap{I}\kern 1.7pt R}}}
\def\Z{\ifmmode\Zmath\else$\Zmath$\fi}
\def\Q{\ifmmode\Qmath\else$\Qmath$\fi}
\def\N{\ifmmode\Nmath\else$\Nmath$\fi}
\def\C{\ifmmode\Cmath\else$\Cmath$\fi}
\def\R{\ifmmode\Rmath\else$\Rmath$\fi}

\def\barray{\begin{eqnarray}}
\def\earray{\end{eqnarray}}
\def\beq{\begin{equation}}
\def\eeq{\end{equation}}

\def\no{\noindent}

\def\gpar{g_\parallel}
\def\gperp{g_\perp}

\def\Jb{\bar{J}}
\def\dx{\frac{d^2 x}{2\pi}}

\def\rap{\beta}
\def\s{\sigma}
\def\spec{\zeta}
\def\comb{\frac{\rap\theta}{2\pi} }
\def\Ga{\Gamma}

\def\L{{\cal L}}
\def\g{{\bf g}}
\def\K{{\cal K}}
\def\I{{\cal I}}
\def\M{{\cal M}}
\def\F{{\cal F}}

\def\gpar{g_\parallel}
\def\gperp{g_\perp}
\def\Jb{\bar{J}}
\def\dx{\frac{d^2 x}{2\pi}}
\def\imag{\Im {\it m}}
\def\real{\Re {\it e}}
\def\Jbar{{\bar{J}}}
\def\kh{{\hat{k}}}
\def\Im{{\rm Im}}
\def\Re{{\rm Re}}
\def\ti{{\tilde{\phi}}}
\def\tR{{\tilde{R}}}
\def\tS{{\tilde{S}}}
\def\tF{{\tilde{\cal F}}}
\def\ba{\bar{a}}
\def\bb{\bar{b}}
\def\be{\bar{\vep_0}}
\def\bD{\bar{\Delta_0}}

\newenvironment{fig}{\linespread{1.0} \begin{figure}}{\end{figure}%
\linespread{1.3}}
\newcommand{\fl}{\hspace*{-\mathindent}}
\newcommand\phup{^{\phantom p}}
\newtheorem{definition}{Definition}[section]
\newtheorem{lemma}{Lemma}[section]
\newtheorem{theorem}{Theorem}[section]
\begin{center}
{\Large\bf  BEC-BCS crossover in a
$p+ip$-wave pairing Hamiltonian coupled to bosonic molecular pairs}\\
~~\\

 {\large Clare Dunning$^{(1)}$, Phillip S. Isaac$^{(2)}$, Jon Links$^{(2)}$,
\\ 
 Shao-You Zhao$^{(2)}$.}\\[3mm]
{\em $^{(1)}$School of Mathematics, Statistics and Actuarial
Science,
The University of Kent, CT2 7NZ, UK,}\\
{\em $^{(2)}$Centre for Mathematical Physics, School of Mathematics
and Physics, The University of Queensland 4072,
 Australia}
\end{center}

\begin{abstract}
We analyse a $p+ip$-wave pairing BCS Hamiltonian, coupled to a single bosonic degree
of freedom representing a molecular condensate, and investigate the nature of the BEC-BCS
crossover for this system. For a suitable restriction on the coupling parameters, we show
that the model is integrable and we derive the exact solution by the algebraic Bethe
ansatz. In this manner we also obtain explicit formulae for correlation functions and compute these for several cases. 
We find that the
crossover between the BEC state and the strong pairing $p+ip$ phase is smooth for this
model, with no intermediate quantum phase transition.
\end{abstract}


\begin{tableofcontents}
\end{tableofcontents}

\section{Introduction}

Progress in cold atom physics has yielded many studies into the nature of the BEC-BCS crossover \cite{becbcs}. Early theoretical accounts emphasized the need to study Hamiltonians which explicitly incorporate coupling between Cooper pairs of atoms and bosonic molecular modes \cite{hkcw01}. Several works extended this approach to the case of $p$-wave paired systems \cite{o05}, a scenario that is experimentally accessible \cite{gsbj07}. Currently there is substantial interest in $p+ip$-wave paired systems \cite{pip}, which has been primarily motivated by the seminal work of Read and Green \cite{rg00} who illustrated the topological distinctions of the quantum phases occuring in this setting. Our objective here is to study a $p+ip$-wave pairing Hamitonian which is coupled to a bosonic molecular degree of freedom to investigate the BEC-BCS crossover in this context. Our approach is to employ exact Bethe ansatz methods for the analysis. 

There have been many exact analyses of the $s$-wave pairing reduced BCS Hamiltonian using the solution provided by Richardson \cite{r63}. These works were particularly prevalent in the wake of experiments conducted on metallic nanograins \cite{vr01}. A comprehensive understanding of the model's mathematical property of {\it integrability} has been developed \cite{crs97} which has lead in particular to some in-depth investigations through the use of exact computation of correlation functions \cite{ao02}. There have been efforts to extend these integrable methods to investigate models where there is coupling between Cooper pairs and bosonic molecular modes \cite{ddep04}. Generally, these examples fall into a class of generalised Dicke/Tavis-Cummings type integrable models \cite{gaudin}. They have the shortcoming that the pair-pair scattering terms found in the Hamiltonians of \cite{hkcw01} are not present, with only pair-molecule scattering terms appearing. 

More recently it has been established that an integrable model also exists for $p+ip$-wave pairing \cite{ilsz09,s09,dilsz10,rdo10}. Integrability in this instance stems from a trigonometric solution of the classical Yang--Baxter equation, in contrast to the rational solution associated with the integrable $s$-wave case. We will show below that an extension of this model through coupling to a bosonic degree of freedom, whilst maintaining pair-pair scattering interactions, is integrable for some restriction of the coupling parameter space. We will derive the exact solution of the Hamiltonian's energy spectrum and certain correlation functions and use these results to study the BEC-BCS crossover.

This paper is organized as follows. We begin Section 2 by introducing a general Hamiltonian describing a $p+ip$-wave pairing BCS model coupled to a bosonic molecular degree of freedom. 
Subsection 2.1 discusses the limiting case of the uncoupled system, in which the extreme limits of BEC and strong pairing BCS ground states are found. Subsection 2.2 establishes suitable constraints on the Hamiltonian's coupling parameters for which the system is integrable, while subsection 2.3 develops the exact solution via algebraic Bethe ansatz methods. The ground-state root structure of the Bethe ansatz equations is determined in subsection 2.4, and based on these results it is shown in 2.5 that the ground-state wavefunction topology is trivial so no topological phase transition exists in the integrable case. Since the integrable case connects the extreme BEC and strong pairing BCS ground states, these belong to the same quantum phase. Section 3 is devoted to the study of correlation function. Subsection 3.1 deals with one-point correlation functions and particular attention is given to the boson fraction expectation value. Subsection 3.2 deals with two-point functions and the boson-Cooper pair fluctuations are studied in some depth. Conclusions are summarised in Section 4. An Appendix on a mean-field treatment of the model is also included.

\section{Model Hamiltonian}

We consider a 2-dimensional $p+ip$-wave pairing BCS model coupled to a single
bosonic degree of freedom where the Hamiltonian of the model is
\begin{eqnarray}
H&=&\delta b^\dag b+\sum_{\bf k}{{\bf k}^2\over 2m}c^\dag_{\bf k}
     c_{\bf k}
 -{G\over 4}\sum_{{\bf k}\ne \pm{\bf k}'}(k_x-ik_y)(k'_x+ik'_y)
     c^\dag_{\bf k}c^\dag_{-{\bf k}}c_{\bf k'}c_{-{\bf k}'}
 \nonumber\\ &&\mbox{}
 -{K\over 2}\sum_{\bf k}\left((k_x-ik_y)
     c^\dag_{\bf k}c^\dag_{-{\bf k}}b+{\rm h.c.}\right).
     \label{de:Hamiltonian}
\end{eqnarray}
One sees that when $\delta=K=0$, the Hamiltonian becomes the 
integrable $p+ip$ pairing BCS model \cite{ilsz09} with  $c_{\bf k}$ and
$c^\dag_{\bf k}$ being destruction and creation operators of
2-dimensional polarised fermions,  ${\bf k}$ and  $m$ the momentum
and mass of the fermions and $G$ a coupling constant which is
positive for an attractive $p+ip$ interaction. In the above Hamiltonian,
the bosonic mode with destruction and creation operators $b$,
$b^\dag$ is associated to a zero-momentum molecular condensate. The
interconversion between Cooper pairs and molecules is controlled by
the coupling $K$. The sign of $K$ is not important since it can be
changed by the unitary transformation $b\rightarrow -b$. Included in
the Hamiltonian is the detuning $\delta$ which accounts for the energy splitting by a 
magnetic field due to the difference between the magnetic
moment of the molecules and that of the Cooper pairs. Hereafter we
set $m = 1$. This model is integrable if we set
$\delta=-F^2G, K=FG$ with $F$ being a free variable, which will be proved below. Before considering that, 
it is useful to first examine the ground-state phases of the uncoupled system.

\subsection{Limiting case of the uncoupled system}

Setting $\delta=K=0$ the Hamiltonian (\ref{de:Hamiltonian}),
restricted to the Hilbert subspace where the bosonic degree of
freedom is in the vacuum state, is the $p+ip$ model. For the extended
model (\ref{de:Hamiltonian}), with $\delta=K=0$ on the full Hilbert
space, the ground state of the system is of the form
\begin{eqnarray}
|\phi\rangle = |\psi_{BCS}\rangle \otimes |N_b\rangle
\label{unstate}
\end{eqnarray}
where $|\psi_{BCS}\rangle$ is a ground state associated with the
$p+ip$ Hamiltonian and $|N_b\rangle$ is a bosonic number state. For the ground state we
need to consider the optimal choice of the boson number $N_b$ which
yields the lowest energy. Since the detuning is zero in this limit,
the ground state will be one which provides the mimimum energy of
$|\psi_{BCS}\rangle$ with respect to variations of the Cooper pair
number.

To elucidate the ground state structure in this limit we recall
results from \cite{dilsz10} for the $p+ip$ model, which has three
ground-state phases called weak coupling, weak pairing, and strong
pairing. Letting $N_C$ denote the number of Cooper pairs, we set
$x_C=N_C/{\cal L}$ as the filling fraction, and $g=G{\cal L}$. Throughout, 
$2{\cal L}$ denotes the total number of momentum levels such that ${\cal L}$ is the
number of momentum pairs. The
three phases are characterized by the constraints shown in Table 1.
In the weak coupling phase the ground-state energy is positive, on
the Moore-Read line it is zero, and in all other cases it is
negative. Ground states in the weak pairing and strong pairing
phases, with filling fractions $x_C^W$, $x_C^S$, are dual whenever $
x_C^W + x_C^S = 1 -g^{-1}, $ with the two ground states having the
same energy. The Read-Green state is self-dual. The Read-Green
condition $x_C =  (1 - {g^{-1}})/2$ gives the state with the lowest
possible energy, for all $g>1$, with respect to variations in $x_C$.
For $g<1$, corresponding to the weak coupling phase, the lowest
possible energy is given by the vacuum since all ground states with
$x_C>0$ have positive energy in this phase. The only phase for which
the ground-state wavefunction is topologically non-trivial is the
weak pairing phase \cite{dilsz10}.

\begin{center}
\begin{tabular}{|c|c|}
\hline
Phase &   Filling fraction $x_C$ \\
\hline \hline weak coupling
& $ x_C > 1 - {g^{-1}}$  \\
\hline Moore-Read  line
& $ x_C = 1 - {g^{-1}}$  \\
\hline weak pairing
& $  (1 - {g^{-1}})/2 < x_C < 1 - {g^{-1}} $  \\
\hline Read-Green  line
& $  x_C =  (1 - {g^{-1}})/2$  \\
\hline strong pairing
& $  x_C <  (1 - {g^{-1}})/2$  \\
\hline
\end{tabular}
\vspace{0.1 cm}

Table 1.- Ground-state phases of the $p+ip$ model.
\end{center}

In view of the above we can determine the ground-state structure of
(\ref{de:Hamiltonian}) when $\delta=K=0$. We let $x=x_b+x_C$ denote
the filling fraction of the system, where $x_b=N_b/{\cal L}$. If
$g<1$ all $p+ip$ states with $x_C>0$ have positive energy, so the
ground state is obtained by choosing $x_b=x$ and $x_C=0$, giving a
pure BEC state for (\ref{unstate}) with zero energy. For $g>1$ the $p+ip$ ground
states have negative energy. If $x>(1-g^{-1})/2$ we choose
$x_C=(1-g^{-1})/2$ so the $p+ip$ state is the Read-Green state,
which has the minimum energy with respect to variations of $x_C$.
This then leaves $x_b=x-(1-g^{-1})/2$ so (\ref{unstate}) is mixed.
Finally if $x<(1-g^{-1})/2$ the $p+ip$ state is in the strong
pairing phase. The energy is miminised by choosing $x_b=0$. This
leads to the classification shown in Table 2.

\newpage
\begin{center}
\begin{tabular}{|c|c|c|c|c|c|}
\hline
Phase & $g$&  Filling fraction  $x$& $|\psi_{BCS}\rangle$ & $N_b$ \\
\hline \hline BEC
& $g<1 $ & all   & vacuum & $N$ \\
\hline Mixed
& $g>1$ &$ x>(1-g^{-1})/2$  &  Read-Green & $0<N_b<N$\\
\hline BCS
& $g>1$ & $ x<(1-g^{-1})/2$ &  strong pairing & $0$\\
\hline
\end{tabular}
\vspace{0.1 cm}

Table 2.- Ground-state phases of the Hamiltonian
(\ref{de:Hamiltonian}) for $\delta=K=0$.
\end{center}

To investigate the crossover between the BEC state and the BCS state we
may start with $g>1$ and $K=\delta=0$ in the Hamiltonian
(\ref{de:Hamiltonian}) so the ground state consists of the strong
pairing $p+ip$ state and the bosonic vacuum provided 
$x<(1-g^{-1})/2$. By turning on $K$ and $\delta$ we obtain an
interacting system of Cooper pairs and bosons. Next we vary $g$ such
that $g<1$, and then turn off $K$ and $\delta$. The ground state
will now consist of the $p+ip$ vacuum and a bosonic number state.
The question we ask is whether the system experiences a 
phase transition as we pass from the strong pairing BCS state to the
BEC state in this manner. Importantly, the coupling parameters can be varied 
such that the Hamiltonian remains integrable as we move between the BEC state and the strong pairing BCS state.

\subsection{Integrability conditions for the coupled system}

It is convenient to first perform a transformation on the Hamiltonian
(\ref{de:Hamiltonian}). We enumerate the complex momenta ${\bf k}=k_x+ik_y$, with $k_y$ in
the upper half-plane, by integers $j=1,...,\L$. Implementing the canonical transformation
\begin{eqnarray*}
s_j={k_x-ik_y\over |{\bf k}|}c_{\bf k}c_{-{\bf k}},\quad\quad
 z_j=|{\bf k}|,
\end{eqnarray*}
we may rewrite the Hamiltonian (\ref{de:Hamiltonian}) as
\begin{eqnarray}
H=\delta N_b+(1+G) H_0-GQ^\dag Q-K Q^\dag b -Kb^\dag Q,
\label{de:Hamiltonian-z}
\end{eqnarray}
where we have defined
\begin{eqnarray}
&&N_b=b^\dag b,\qquad\qquad\, {N}_{j}=s_j^\dag s_j, \,\,\qquad j=1,\ldots, {\cal L},\nonumber\\
&&H_0= \sum_{j=1}^{\cal L} z_j^2 N_j, \quad \,\,\,\, Q^\dagger= \sum_{j=1}^{\cal L} z_j s_j^\dagger.
\label{de:transformtwo}
\end{eqnarray}
provided we restrict to the subspace of the Hilbert space which excludes blocked states (see \cite{vr01} for a discussion of the blocking effect). This restriction is sufficient to study the ground-state properties when the total fermion number is even.  We use ${N}$ to stand for the {\it pair number operator} which is the sum of
the boson and pairing number operators; namely, ${N}={N}_b+{N}_c$ with
${N}_c=\sum_{j=1}^{\cal L}{ N}_{j}$. We note that ${N}$ commutes with the Hamiltonian
(\ref{de:Hamiltonian}). This allows us to block diagonalise the Hamiltonian into sectors
labelled by the eigenvalues of $N$, which are non-negative integers. Hereafter we will
adopt the practice to interchangably use the symbol $N$ to denote the pair number operator
and its eigenvalues.

Now we show that for a  suitable restriction on the coupling parameters of
(\ref{de:Hamiltonian}) the model is integrable. The integrable manifold is defined by the relations
\begin{eqnarray}
\delta=-F^2G,\,\qquad K=FG
\label{de:manifold}
\end{eqnarray}
with $F$ being a free variable. Under this constraint, the
Hamiltonian (\ref{de:Hamiltonian}) becomes 
\begin{eqnarray}
H&=&-F^2G N_b+(1+G) H_0-GQ^\dag Q-FG Q^\dag b -FGb^\dag Q.
     \label{de:Hamiltonian-int}
\end{eqnarray}
We will prove the integrability of the above
Hamiltonian by using the Quantum Inverse Scattering Method \cite{Fadeev79}. Our approach is a generalisation of the method detailed in \cite{dilsz10}.

Let $V$ be the 2-dimensional $U_q(sl(2))$-module and $R(\lambda)\in
\mbox{End}(V\otimes V)$ the six-vertex solution of the Yang-Baxter equation
$$
R_{12}(\lambda/\mu)R_{13}(\lambda)R_{23}(\mu)
=R_{23}(\mu)R_{13}(\lambda)R_{12}(\lambda/\mu)
$$
acting on the three-fold space $V\otimes V\otimes V$.
The $R$-matrix, which depends on the spectral parameter
$\lambda$ and the crossing parameter $q$, explicitly reads 
\begin{eqnarray*}
 R(\lambda)=\left(\begin{array}{ccccc}
 \lambda q^2-\lambda^{-1}q^{-2}&0&|&0 & 0\\
 0&\lambda-\lambda^{-1} & |& q^2-q^{-2}&0\\
 -&-& \mbox{}&-&-\\
 0& q^2-q^{-2}&|&\lambda-\lambda^{-1}& 0\\
 0& 0&|&0&\lambda q^2-\lambda^{-1}q^{-2}
 \end{array} \right). 
\end{eqnarray*}
We construct the Yang-Baxter algebra by using the
$R$-matrix and the $L$-operator $L(\lambda)$  through
the Yang-Baxter relation (YBR)
\begin{eqnarray}
R_{12}(\lambda/\mu)L_{1j}(\lambda)L_{2j}(\mu)=
 L_{2j}(\mu)L_{1j}(\lambda)R_{12}(\lambda/\mu). \label{eq:YBR}
\end{eqnarray}
Here $L_{\alpha j}(\lambda)\in \mbox{End}(V\otimes V)$ is a $2\times 2$
matrix of operators. In the framework of quantum integrable
systems, the subscript $\alpha$ labels the auxiliary space,
while entries of the matrix are operators acting on the $j$th
quantum space.

A well-known $L$-operator is realised by the $R$-matrix itself which, using local
creation $s^\dag$ and destruction operators $s$, is expressed as 
\begin{eqnarray*}
L_{\sigma  i}(\lambda)&=&\left(\begin{array}{cc}
 \lambda q^{(2{N}_{i}-1)}-\lambda^{-1}q^{-(2{N}_{i}-1)} &
 (q^2-q^{-2})s_i\\ (q^2-q^{-2}) s_i^\dag &
 \lambda q^{-(2{N}_{i}-1)}-\lambda^{-1} q^{(2{N}_{i}-1)}
 \end{array}\right)_{(\sigma )} \\
\end{eqnarray*}
where ${N}_{i}$ is the local number operator with the definition
${N}_{i}=s_i^\dag s_i$. The operators $s_i^\dag$, $s_i$ and
${N}_{i}$ are generators of the quantum algebra $U_{q}(sl(2))$. In the 2-dimensional representation they satisfy
the relation
\begin{eqnarray*}
[s_i,s_j^\dag]=\delta_{ij}(I-2{N}_{i}).
\end{eqnarray*}
A realisation of the $L$-operator using the
$q$-boson algebra was given by Kundu \cite{Kundu07}:
\begin{eqnarray*}
\tilde L_{\sigma  b}(\lambda)=\left(\begin{array}{cc}
 \lambda q^{2N_b}-i\lambda^{-1}q^{-2 ( N_b+1)} &
 -e^{i\pi/4}(q^4-q^{-4})^{1/2} b \\ -e^{i\pi/4}(q^4-q^{-4})^{1/2} b^\dag  &
 \lambda q^{-2 N_b}+i\lambda^{-1}q^{2( N_b+1)}
 \end{array}\right)_{(\sigma )}.
\end{eqnarray*}
The subscript $b$ in the $L$-operator $\tilde
L_{\sigma  b}(\lambda)$  stands for the bosonic quantum space. The local $q$-boson
operators $b_q$, $b_q^\dag$ and $N_b=b_q^\dag b_q$ have the
following commutation relation
$$
 [b_q,b_q^\dag]={q^{2(2N_{b}+1)}+q^{-2(2N_{b}+1)}\over
q^{2}+q^{-2}}.
$$
It can be seen that when $q\rightarrow 1$, $b_q$ and $b_q^\dag$
become the usual bosonic destruction and creation operators $b$ and
$b^\dag$. With the help of the mapping
\begin{eqnarray*}
\tilde L_{\sigma  b}(\lambda) \rightarrow -e^{-i\pi/4}(q^4-q^{-4})^{1/2}
\mbox{diag}\left(q^{1/2},q^{-1/2}\right)\cdot
 \tilde L_{\sigma  b}(\lambda) \cdot
 \mbox{diag}\left(q^{1/2},q^{-1/2}\right)
\end{eqnarray*}
and the variable shift $\lambda\rightarrow
-e^{-i\pi/4}(q^4-q^{-4})^{1/2} \lambda$ the $L$-operator, which still satisfies (\ref{eq:YBR}), becomes
\begin{eqnarray*}
&&\tilde L_{\sigma  b}(\lambda)=\left(\begin{array}{cc}
   (\tilde L_{\sigma  b})_{11}& (\tilde L_{\sigma  b})_{12}\\
   (\tilde L_{\sigma  b})_{21} &(\tilde L_{\sigma  b})_{22}
 \end{array}\right)_{(\sigma )},
\end{eqnarray*}
where the elements are
\begin{eqnarray*}
&&(\tilde L_{\sigma  b})_{11}=\lambda q^{(2N_b+1)}
  -(q^4-q^{-4})\lambda^{-1}q^{-(2N_b+1)},\\
&&(\tilde L_{\sigma  b})_{12}=(q^4-q^{-4}) b_q,\\
&&(\tilde L_{\sigma  b})_{21}=(q^4-q^{-4}) b^\dag_q,\\
&&(\tilde L_{\sigma  b})_{22}=\lambda q^{-(2N_b+1)}
 +(q^4-q^{-4})\lambda^{-1}q^{(2N_b+1)}.
\end{eqnarray*}

Now we define the monodromy matrix
\begin{eqnarray}
T_\sigma(\lambda)&=&g_\sigma\tilde L_{\sigma b}(\lambda z_b^{-1})L_{\sigma {\cal L}}(\lambda
z_{\cal L}^{-1})\cdots L_{\sigma 2}(\lambda z_{2}^{-1})
L_{\sigma 1}(\lambda z_1^{-1})) \label{de:monodromy}
\end{eqnarray}
with the diagonal matrix $g_\sigma=\mbox{diag} (e^{-i\alpha},e^{i\alpha})_{(\sigma)}$.
Using the YBR (\ref{eq:YBR}), the
following equation holds for the monodromy matrix
$$
R_{\sigma\rho}(\lambda/\mu)T_\sigma(\lambda)T_\rho(\mu)=
 T_\rho(\mu)T_\sigma(\lambda)R_{\sigma\rho}(\lambda/\mu). 
$$
This relation ensures the
commutation relation
$$
[t(\lambda),t(\mu)]=0,\quad \forall\,\, \lambda,\mu
$$
where $t(\lambda)$ is the transfer matrix defined by $t(\lambda)={\rm
tr}_\sigma \left[T_\sigma(\lambda)\right]$.

Expanding the transfer matrix $t(\lambda)$ in orders of the spectral
parameter $\lambda$
$$
t(\lambda)=\sum_{i=-{\cal L}-1}^{{\cal L}+1}t^{(i)} \lambda^i,
$$
we find that the coefficients commute with each other
$$
[t^{(i)},t^{(j)}]=0
$$
for all $i,j$. In this manner we may construct an
integrable system by using the coefficients $t^{(i)}$. The leading
terms of the expansion are
\begin{eqnarray*}
t^{({\cal L}+1)}&=&\left(z_b^{-1}\prod_{i=1}^{\cal L}z_i^{-1}\right)
 \left(
 e^{-i\alpha}q^{2{N}-{\cal L}+1} + h.c.\right),\nonumber\\
t^{({\cal L})}&=&0,\nonumber\\
t^{({\cal L}-1)}&=&-\left(z_b^{-1}\prod_{i=1}^{\cal
L}z_i^{-1}\right)
 \sum_{j=1}^{\cal L}z_j^2\left(
  e^{-i\alpha}q^{2{N}-{\cal L}+1}q^{-(4{N}_{j}-2)} +h.c.\right)\nonumber\\
 &&\mbox{}-z_b^2(q^4-q^{-4})\left(z_b^{-1}\prod_{i=1}^{\cal L}z_i^{-1}\right)
 \left( e^{-i\alpha}q^{2{N}-{\cal L}+1} q^{-(4N_b+2)}-
 h.c\right)
 \nonumber\\
&&\mbox{}+
 (q^2-q^{-2})^2\left(z_b^{-1}\prod_{i=1}^{\cal L}z_i^{-1}\right)
  \nonumber\\ &&\quad\quad\times
 \sum_{j<k}^{\cal L}z_jz_k %
 \left(
  e^{-i\alpha}q^{2{N}-{\cal L}+1}\prod_{l=j+1}^{k-1} q^{-(4{N}_{l}-2)} s_ks_j^\dag +
  h.c.\right)\nonumber\\
&&\mbox{}+
 (q^2-q^{-2})^{2}(q^2+q^{-2})\left(z_b^{-1}\prod_{i=1}^{\cal L}z_i^{-1}\right)
\nonumber\\ &&\quad\quad\times
 \sum_{j=1}^{\cal L}z_bz_j
 \left(
 e^{-i\alpha}q^{2{N}-{\cal L}+1}q^{-2 N_b-1}
 \prod_{l=j+1}^{\cal L} q^{-(4{N}_{l}-2)} b_qs_j^\dag +h.c. \right).
\end{eqnarray*}
Introducing the notation
$$
q=e^{i\beta},\quad \beta=\eta p,\quad\alpha-\beta(2{N}-{\cal
L}+1)= {\eta t},
$$
we define a Hamiltonian $\tilde{H}$ by using the coefficient
$t^{({\cal L}-1)}$:
\begin{eqnarray}
\tilde{H}&=&(q^2-q^{-2})^{-2}z_b\prod_{i=1}^{\cal L}z_i\,\,t^{({\cal L}-1)}
  \nonumber\\
 &=&-{1\over \sin^2(2\eta p)}\sum_{j=1}^{\cal L}z_j^2
 \sin^2\left(\eta{t+4p{N}_{j}-2p\over 2}\right)
 +{1\over 2\sin^2(2\eta p)}\sum_{j=1}^{\cal L}z_j^2\nonumber\\
 &&\mbox{}+2 z_b^2\cot(2\eta p)\sin(\eta(t+4pN_b+2p))
 \nonumber\\
&&\mbox{}+\sum_{j<k}^{\cal L}z_jz_k
 \left(
 e^{-i\eta t}\prod_{l=j+1}^{k-1} e^{-i\eta p(4{N}_{l}-2)} s_ks_j^\dag +
 e^{i\eta t}\prod_{l=j+1}^{k-1} e^{i\eta p(4{N}_{l}-2)} s_k^\dag s_j \right)\nonumber\\
&&\mbox{}+2\cos(2\eta p)\sum_{j=1}^{\cal L}z_bz_j
  \left( e^{-i\eta(t+2pN_b+p)}
 \prod_{l=j+1}^{\cal L} e^{-i\eta(4{N}_{l}-2)} b_q s_j^\dag\right.\nonumber\\
&&\mbox{}\quad\quad +\left.
  e^{i\eta(t+2pN_b+p)}
 \prod_{l=j+1}^{\cal L} e^{i\eta(4{N}_{l}-2)} b^\dag_qs_j  \right).
 \label{eq:Hamiltonian-gen}
\end{eqnarray}
Let $G=2p/t$ and $F=2z_b$. Taking the limit $\eta\rightarrow 0$,
 we obtain the following Hamiltonian 
\begin{eqnarray}
H&=&-G \lim_{\eta\rightarrow 0} \left(\tilde{H}
           - {1\over 2\sin^2(2\eta p)}\sum_{j=1}^{\cal L}z_j^2 +\sum_{j=1}^{\cal L}z_j^2{(t-2p)^2\over16p^2}-{z_b^2(t+2p)\over p}\right)\nonumber\\
   &=&\sum_{j=1}^{\cal L}z_j^2{N}_{j}-{F^2G}N_b
 -G\sum_{j<k}^{\cal L}z_jz_k
 \left(s_ks_j^\dag +h.c\right)
 -{FG}\sum_{j=1}^{\cal L}z_j \left(bs_j^\dag + h.c
 \right).\nonumber\\
  \label{de:Hamiltonian-p}
\end{eqnarray}
Utilizing (\ref{de:transformtwo}) we find that (\ref{de:Hamiltonian-p}) is equivalent to
(\ref{de:Hamiltonian-int}). Therefore we have established that the constraint
(\ref{de:manifold}) defines an integrable manifold in the coupling parameter space of
(\ref{de:Hamiltonian}).


\subsection{Algebraic Bethe ansatz solution}

The eigenvalues of the Hamiltonian (\ref{de:Hamiltonian-p}) can be
obtained by using the algebraic Bethe ansatz. Again, we follow the procedure of \cite{dilsz10} and only present the main results.  Rewriting the
monodromy matrix $T_\sigma (\lambda)$ (\ref{de:monodromy}) by using 
global quantum operators $A(\lambda),B(\lambda),C(\lambda)$ and
$D(\lambda)$ defined by 
\begin{eqnarray*}
T(\lambda)=\left(\begin{array}{cc}
 A(\lambda)&B(\lambda)\\
 C(\lambda)&D(\lambda)\end{array}\right)_{(\sigma)},
\end{eqnarray*}
the transfer matrix $t(\lambda)$ becomes
$$
t(\lambda)={\rm tr}_\sigma \left[T_\sigma (\lambda)\right]=A(\lambda)+D(\lambda).
$$
The Bethe states of the system are defined by
$$
|\Phi(\{\mu\})\rangle =\prod_{i=1}^{N} C(\mu_i)|0\rangle,
$$
where $|0\rangle$ is the vacuum state with the definition
$$
b|0\rangle= s_i|0\rangle=0
$$
for all $i= 1,\ldots,{\cal L}.$

By using the standard algebraic Bethe ansatz method \cite{Fadeev79}, the
eigenvalues of the Hamiltonian (\ref{eq:Hamiltonian-gen})
are given by
\begin{eqnarray*}
\tilde{E} &=&2z_b^2\cot(2\eta p)\sin(\eta(t+2p))
 -\left({\sin^2(\eta(-t+2p)/2)\over \sin^2(2\eta p)}\right)
  \sum_{i=1}^{\cal L}z_i^2 \nonumber\\ &&\mbox{}
 +{1\over 2\sin^2(2\eta p)}\sum_{i=1}^{\cal L}z_i^2
 -{\sin(\eta(t+2p))\over\sin(2\eta p)}\sum_{j=1}^{N}\mu_j^2.
\end{eqnarray*}
Here, the parameters $\mu_j$ $(j=1,2,\ldots,{N})$ satisfy the following
Bethe ansatz equations
\begin{eqnarray*}
&& e^{-2i\alpha}{\mu_j z_b^{-1}q -(q^4-q^{-4})\mu_j^{-1}z_b q^{-1}
\over
  \mu_j z_b^{-1} q^{-1}+(q^4-q^{-4})\mu_j^{-1}z_b q}
 \prod_{i=1}^{\cal L}
 {\mu_j z_i^{-1}q^{-1}-\mu_j^{-1}z_iq\over
  \mu_j z_i^{-1}q-\mu_j^{-1}z_iq^{-1}}
 \nonumber\\
 &&
 = \prod_{k\ne j}^{N}
  {\mu_j\mu_k^{-1}q^{-2}-\mu_j^{-1}\mu_k q^2\over
   \mu_j\mu_k^{-1}q^2-\mu_j^{-1}\mu_k q^{-2}}\quad.
\end{eqnarray*}

Taking the limit $\eta\rightarrow 0$, we obtain the eigenvalues of
(\ref{de:Hamiltonian-p}):
$$
E =(1+G)\sum_{j=1}^{N}\mu_j^2 
$$
subject to the Bethe ansatz equations 
$$
{G^{-1}+2{N}-{\cal L}-1\over \mu_j^2}
  +{4z_b^2\over \mu_j^4}
 +\sum_{i=1}^{\cal L}{1\over\mu_j^2-z_i^2 }
 =\sum_{k\ne j}^{N}{2\over\mu_j^2-\mu_k^2}
$$
for $j=1,\ldots,{N}$.
For convenience, throughout the remainder of the paper we will
simplify notation by making the substitutions
$$
\lambda_j^2\mapsto\lambda_j,\quad\quad \mu_j^2\mapsto\mu_j,
$$
such that the Bethe ansatz equations take the form
\begin{eqnarray}
{G^{-1}+2{N}-{\cal L}-1\over \mu_j}
  +{4z_b^2\over \mu_j^2}
 +\sum_{i=1}^{\cal L}{1\over\mu_j-z_i^2 }
 =\sum_{k\ne j}^{N}{2\over\mu_j-\mu_k}.\quad\quad
  \label{BAE}
\end{eqnarray}
In the $\eta\rightarrow 0$ limit the Bethe states and their dual states are defined by 
\begin{eqnarray}
&&|\phi(\{\lambda\})\rangle=\prod_{j=1}^{N}
       {\cal C}(\lambda_j)|0\rangle,\label{de:BS}\\
&&\langle\phi(\{\mu\})|=\langle0|\prod_{j=1}^{N}
 {\cal B}(\mu_j),
\nonumber
\end{eqnarray}
where ${\cal C}$ and ${\cal B}$ are the global creation and destruction operators given by
\begin{eqnarray}
{\cal C}(\lambda)
  &=& {2z_b b^\dag\over \lambda}+
 \sum_{j=1}^{\cal L}{z_j s_j^\dag\over\lambda-z_j^2},\label{de:operator-C}\\
{\cal B}(\mu)
  &=& {2z_b b\over \mu}+
 \sum_{j=1}^{\cal L}{z_j s_j\over\mu-z_j^2}\label{de:operator-B}
\end{eqnarray}

\subsection{Ground-state root structure} \label{tl}

It is necessary to understand the character of the roots of
(\ref{BAE}) which correspond to the ground state of the model. By
adding an appropriate constant term to (\ref{de:Hamiltonian-int}),
all matrix elements are real and negative on each sector with fixed
${N}$. From numerical studies of (\ref{BAE}) we find solutions for
which all the roots $\lambda_j$ are real and negative. From
(\ref{de:operator-C}), and an appropriate rescaling ${\cal
C}(\lambda) \rightarrow - {\cal C}(\lambda)$, we see that these
roots give rise to an eigenvector with positive components. This
eigenvector necessarily corresponds to the ground state as a result
of the Perron-Frobenius theorem. This theorem also tells us that
there is a unique solution set with the property that all
$\lambda_j$ are real and negative.


While we are unable to prove existence of a solution set with the property of all roots being real and negative in a general finite system, we can establish existence of such a set in the thermodyanamic limit.   
To analyze the thermodynamic limit of the model,
${\cal L} \rightarrow \infty$,
$N \rightarrow \infty$ such that the filling fraction $x=N/{\cal L}$
remains finite,  we
follow the approach of \cite{dilsz10,rdo10} used to treat the strong pairing phase of the $p+ip$ model. Making use of the
following notations
\begin{eqnarray*}
 g=G{\cal L},\quad f={2z_b\over \sqrt{\cal L}},\quad
 q={G^{-1}+2N-{\cal L}-1\over {\cal L}}
\end{eqnarray*}
and assuming that the ground-state roots $\mu_j$
become dense on an interval $[a,b]$ of the negative real axis, the
BAEs (\ref{BAE}) become the integral equation
\begin{equation} 
\int_0^\omega d \vep \frac{ \rho(\vep)}{ \vep - \mu} -
\frac{q}{\mu}-\frac{f^2}{\mu^2} - P \int_a^b d\mu' \frac{2r(\mu')}{ \mu'- \mu} =
0 \label{19}
\end{equation}
where $r(\mu)$ is the density of the roots, and $\rho(\vep)$ is the
density of the $\vep=z^2$ located on the positive
real axis such that
$$\int_0^\omega d\vep\, \rho(\vep)=1. $$
The filling fraction $x$ and intensive energy $\displaystyle e_0=\lim_{{\cal
L}\rightarrow\infty} E_0/{\cal L}$ are given by
\begin{equation} 
x = \int_0^\omega d\mu\,  r(\mu), \;\quad e_0 = \int_0^\omega d\mu \,
\mu \, r(\mu). 
\label{xe} 
\end{equation}
Using standard techniques of complex
analysis, the solution $r(\mu)$ of (\ref{19}) is
\begin{eqnarray*}
r(\mu)&  = &  \frac{R(\mu)}{\pi i} \left[\int_0^\omega d \vep
\frac{\rho(\vep)}{  (\vep-\mu)R(\vep)} - \frac{S}{\mu}-\frac{T}{\mu^2}
\right],
\nonumber \\
R(\mu) &  = &  \sqrt{ (\mu-a)(\mu-b)}, \nonumber\\
S&=& \frac{1}{2\sqrt{ab}}\left(q+\frac{f^2(a+b)}{4ab}\right), \\
T&=& \frac{f^2}{2\sqrt{ab}}
\end{eqnarray*}
with the constraint
\begin{eqnarray}
 q+\frac{f^2(a+b)}{2ab}&=& -\sqrt{ab}\int^\omega_0 d\vep \, \frac{\rho(\vep)}{R(\vep)}.
\label{chempot}
\end{eqnarray}
Evaluating (\ref{xe}) gives
\begin{eqnarray}
\frac{1}{g}&=&\frac{f^2}{\sqrt{ab}}+\int^\omega_0 d\vep\,\frac{\vep\rho(\vep)}{R(\vep)}, \label{gap} \\
e_0&=&f\left(\frac{1}{2}-\frac{a+b}{4\sqrt{ab}}\right) +\frac{1}{2}\int^\omega_0 d\vep\, \vep
\rho(\vep)\left(1-\frac{2\vep-a-b}{2R(\vep)} \right). \label{intnrg}
\end{eqnarray}
The value for $e_0$ is obtained by solving equations
(\ref{chempot},\ref{gap}) for $a,\,b$, and substituting these values
into (\ref{intnrg}). 
Equations (\ref{chempot})-(\ref{intnrg}) are in agreement with mean-field results given in the Appendix.

\subsection{Topology of the ground-state wavefunction}

In the case of the $p+ip$ model the ground-state phases as depicted
in Table 1 are independent of both the distribution of the momentum
variables and the cut-off $\omega$, which is a consequence of the
topological nature of the phases. For the analysis of
(\ref{de:Hamiltonian}) we consider that the momenta  are fixed, so
the parameter space of (\ref{de:Hamiltonian}) is three-dimensional
with $G,\,K,\,\delta$ as the variable coupling constants. For the
two-dimensional surface within the parameter space for which the Hamiltonian
(\ref{de:Hamiltonian}) admits an exact Bethe ansatz solution, 
the ground-state roots are real and negative. We can use this
property to show that the ground-state wavefunction is topologically
trivial in the exactly solvable case. To this end we will adopt the
winding number approach used in \cite{ilsz09,dilsz10} for the $p+ip$
model.

The topological structure of a complex function $\varphi({\bf k})
=\varphi_x({\bf k}) + i \varphi_y({\bf k})$
can be characterized by a winding number $w$. Consideration of the
stereographic projection of the $\bk$-space domain of $\varphi({\bf
k})$, and the stereographic projection of the image of $\varphi({\bf
k})$, induces a map between Riemann spheres $\tilde{\varphi}:
S^2\rightarrow S^2$. We adopt the convention for the stereographic
projections that the point at infinity for both $\bk$-space and the
image of $\varphi({\bf k})$ is associated with the north pole of the
spheres.   The winding number associated with $\tilde{\varphi}$ is
$$
w =  \frac{1}{\pi} \int_{\mathbb{R}^2}  d k_x \; d k_y \frac{
\partial_{k_x} \varphi_x \partial_{k_y} \varphi_y -
\partial_{k_y} \varphi_x \partial_{k_x} \varphi_y }{
(1 + \varphi_x^2 + \varphi_y^2)^2}. 
$$
The key point to
recognise is that a non-zero value of $w$ can only occur if the
north pole is in the image of $\tilde{\varphi}$, which is equivalent
to the statement that $\varphi({\bf k})$ is divergent for some
$\bk$. These concepts generalise to multivariate functions
$\varphi({\bk_1},...\bk_M)$.

Now we turn to the ground state as given by (\ref{de:BS}) and
consider the expansion
\begin{eqnarray*}
|\phi\rangle = \sum_{j=0}^N |\psi_j\rangle \otimes |N_b=N-j\rangle
\end{eqnarray*}
where each $|\psi_j\rangle$ is a state of $j$ Cooper pairs
expressible as
\begin{eqnarray*}
|\psi_j\rangle = \sum_{\bk_1,...,\bk_j} \psi_j(\bk_1,...,\bk_j)
c^\dagger_{\bk_1} c^\dagger_{-\bk_1}...c^\dagger_{\bk_j}
c^\dagger_{-\bk_j}|0\rangle.
\end{eqnarray*}
The possible pole structure of $\psi_j(\bk_1,...,\bk_j)$ can be
deduced from the co-efficient terms of each ${\cal C}(\mu_j)$,
viz. $\gamma(\bk)$ given by $\gamma(\bk)=(k_x-ik_y)/(\mu-{\bf
k}^2)$. Since the $\mu_j$ are all real and negative for the
ground-state these terms do not diverge for any $\bk$.  This
situation should be contrasted with the $p+ip$ model where changes
in the topology of the ground state occur exactly when some of the
roots $\mu_j$ vanish, in which case $\gamma(\bk)$ diverges at
$\bk=0$ \cite{ilsz09,dilsz10}. Hence the functions $\psi_j(\bk_1,...,\bk_j)$
are topologically trivial. This leads to the conclusion that there
is no topological phase transition in the exactly solvable case.  As the
exactly solvable case allows us to crossover from the strong pairing BCS
state to  the BEC state, these two states belong to the same topological phase of 
the ground-state phase diagram.

\section{Correlation functions}

Our conclusion that the crossover between the BCS and BEC ground states
is smooth should manifest in the correlation
functions of the Hamiltonian. Within the framework of the algebraic Bethe ansatz, these may be computed exactly in terms of determinants of matrices whose entries are functions of the roots of (\ref{BAE}). Following the calculations of \cite{dilsz10} (see Appendix A.2), based on results of Slavnov
\cite{Slavnov89}, if the parameters $\mu_i$
satisfy the Bethe ansatz equations (\ref{BAE}), the
scalar products of states for arbitrary parameters $\lambda_j$
are 
\begin{eqnarray}
{\cal S}(\{\mu\},\{\lambda\})&=&\langle\Phi(\{\mu\})|\Phi(\{\lambda\})
   \rangle\nonumber\\
&=&{1\over \prod_{k<l}^{N}
(\lambda_k-\lambda_l)(\mu_l-\mu_k)}
\mbox{det}{\cal G} (\{\mu\},\{\lambda\})\label{eq:sclar-product}
\end{eqnarray}
with
\begin{eqnarray}
{\cal G}_{ij}(\{\mu\},\{\lambda\}) 
&=&\prod_{l\ne i}^{{N} }
 (\mu_l-\lambda_{j})
 \left({4z_b^2\over\lambda_j\mu_i}+
  \sum_{k=1}^{\cal L}{z_k^2\over (\mu_i-z_k^2)(\lambda_j-z_k^2)}
               +\sum_{n\ne i}^{N}{2\mu_n\over(\mu_n-\mu_i)(\lambda_j-\mu_n) }\right).
\nonumber\\
&& \label{eq:matrix-sp}
\end{eqnarray}
Of particular interest is the case when $\mu_i=\lambda_i$ $\forall i$, whereby
\begin{equation}
{\cal S}(\{\lambda\},\{\lambda\}) = \mbox{det}\ {\tilde G}(\{\lambda\})
\label{eq:sclr-pr-eq}
\end{equation}
with
\begin{eqnarray*}
{\tilde G}(\{\lambda\})_{ii} & = & \frac{4z_b^2}{\lambda_i^2}
+\sum_{k=1}^{\cal L}\frac{z_k^2}{(\lambda_i-z_k^2)^2}
-\sum_{n\neq i}^N\frac{2\lambda_n}{(\lambda_i-\lambda_n)^2},\nonumber\\
{\tilde G}(\{\lambda\})_{ij} & = &
\frac{2\lambda_j}{(\lambda_j-\lambda_i)^2},\qquad i\neq j.
\end{eqnarray*}
Equipped with this result, we now proceed to calculate several forms of correlation functions.

In general
an $m$-point correlation function is defined by
\begin{eqnarray*}
F(\{\mu\},\epsilon^1_{i_1},\ldots,\epsilon^m_{i_m},
  \{\lambda\})
 =\langle \phi(\{\mu\})|\epsilon^1_{i_1}\ldots\epsilon^m_{i_m}
  |\phi(\{\lambda\})\rangle,
\end{eqnarray*}
where $\epsilon^j_{i_j}$ stand for any local pairing operators
$s_{i_j},s^{\dag}_{i_j}$, $N_{i_j}$ or bosonic operators
$b,b^{\dag}$ and $N_b$, and the lower indices $i_j$ indicate the
positions of the operators.
With the help of the definition of the global creation and
destruction operators (\ref{de:operator-C},\ref{de:operator-B}), we
may solve the inverse problem for the local operators through
\begin{eqnarray}
s^\dag_j&=&\lim_{v\rightarrow z_j^2}{ v-z_j^2\over z_j}{\cal
C}(v),\quad
s_j=\lim_{v\rightarrow z_j^2}{ v-z_j^2\over z_j}{\cal B}(v),\nonumber\\
b ^\dag&=&\lim_{v\rightarrow 0}{ v\over 2z_b}{\cal C}(v),\quad\quad\quad b
=\lim_{v\rightarrow 0}{ v\over 2z_b}{\cal B}(v). \label{de:inverse}
\end{eqnarray}
Below, instead of representing the local number
operators $N_b$ and ${N}_{m}$ in terms of global operators, we will use the
following commutation relations to compute their correlation
functions
\begin{eqnarray}
&& [N_b,{\cal C}(\lambda)]={2z_b\over \lambda}\, b ^\dag,
 \label{eq:comm-NaC}\\
&&[{N}_{m},{\cal C}(\lambda)]={z_m\over \lambda-z_m^2}\, s_m^\dag.
 \label{eq:comm-NC}
\end{eqnarray}
Note that throughout we assume that the
parameters $\{\mu_i\}$ satisfy the Bethe ansatz equations
(\ref{BAE}). For the off-diagonal one-point functions below, the cardinality of the set 
$\{\mu_i\}$ is one greater than that of $\{\lambda_j\}$, while in all other instances 
they have equal cardinality.  

\subsection{One-point correlation functions}
\begin{itemize}
\item The off-diagonal one-point function for the fermion pair creation operator $s^\dag_m$ is
$$
F(\{\mu\},s^\dag_m,\{\lambda\}) =\langle
\phi(\{\mu\})|s^\dag_m|\phi(\{\lambda\}) \rangle .
$$
\end{itemize}
Substituting the representation of $s_m^\dag$ (\ref{de:inverse}) into
the above definition, we have
\begin{eqnarray*}
 F(\{\mu\},s^\dag_m,\{\lambda\})
=\lim_{v\rightarrow z_m^2}{v-z_m^2\over z_m}
   \langle \phi(\{\mu\})|{\cal C}(\lambda_1)\ldots
   {\cal C}(\lambda_{{N}-1}){\cal C}(v)|0\rangle.
\end{eqnarray*}
It is seen that the one-point function is a limit of the scalar
product (\ref{eq:sclar-product}). Substituting the scalar product
into the above formula, we obtain
$$
F(\{\mu\},s^\dag_m,\{\lambda\})
={\prod_{i=1}^{{N}}(\mu_i-z_m^2)\over \prod_{i=1}^{{N}-1}
(\lambda_i-z_m^2)}
   {\mbox{det}({\cal T}_m^N(\{\mu\},\{\lambda\}) )
\over
   \prod_{k<l}^{{N}-1}(\lambda_k-\lambda_l)\prod_{k<l}^{{N}}(\mu_l-\mu_k) }
$$
with the elements of the ${N}\times {N}$ matrix ${\cal T}_m^N$ given by 
\begin{eqnarray*}
\left({\cal T}_m^N(\{\mu\},\{\lambda\})   \right)_{ij}&=& 
\left({\cal G}(\{\mu\},\{\lambda\})\right)_{ij},
    \qquad j=1,\ldots,{N}-1,\\
\left({\cal T}_m^N(\{\mu\},\{\lambda\})\right)_{i\,{N}}&=&{z_m\over (\mu_i-z_m^2)^2}.
\end{eqnarray*}
We also note that 
$$F(\{\mu\},s^\dag_m,\{\lambda\})
=F(\{\lambda\},s_m,\{\mu\}). $$

\begin{itemize}
\item The off-diagonal one-point function for the boson creation operator $b^\dag$
is 
$$ 
 F(\{\mu\},b^\dag,\{\lambda\})
  =\langle \phi(\{\mu\})|b^\dag|
    \phi(\{\lambda\})\rangle
$$
\end{itemize}
Using a similar method as for the case of $s_m^\dag$, we obtain the
one-point function as follows:
$$
F(\{\mu\},b^\dag,\{\lambda\})
 ={\prod_{i=1}^{{{N}}}\mu_i\over \prod_{i=1}^{{N}-1}\lambda_i}
   {\mbox{det}({\cal T}_b(\{\mu\},\{\lambda\}))\over
   \prod_{k<l}^{{N}-1}(\lambda_k-\lambda_l)\prod_{k<l}^{{{N}}}(\mu_l-\mu_k) }
$$
with 
\begin{eqnarray*}
 ({\cal T}_b(\{\mu\},\{\lambda\}))_{ij}&=& \left({\cal
G}(\{\mu\},\{\lambda\})\right)_{ij},\qquad j=1,\ldots,{N}-1, \\
 ({\cal T}_b(\{\mu\},\{\lambda\}))_{i\,{{N}}}&=&{2z_b\over \mu_i^2}\,\,.
\end{eqnarray*}
Similar to the case above we have 
$$F(\{\mu\},b^\dag,\{\lambda\})
=F(\{\lambda\},b,\{\mu\}). $$

\begin{itemize}
\item To calculate the diagonal one-point function for the Cooper pair number operator
${N}_{m},$ we consider only the functions
$$
F(\{\lambda\},{N}_{m},\{\lambda\})
=\langle\phi(\{\lambda\})|{N}_{m}|\phi(\{\lambda\})\rangle.
$$
\end{itemize}
Using the commutation relation (\ref{eq:comm-NC}) and the fact that 
${N}_{m}|0\rangle=0$, we obtain
\begin{eqnarray}
&&F(\{\lambda\},{N}_{m},\{\lambda\})\nonumber\\ &&
=\sum_{j=1}^{N}{z_m\over \lambda_j-z_m^2}
   \langle\phi(\{\lambda\})|
   {\cal C}(\lambda_1)\ldots  {\cal C}(\lambda_{j-1})s_m^\dag
   {\cal C}(\lambda_{j+1})\ldots{\cal C}(\lambda_N)|0\rangle \nonumber\\
&&=\langle\phi(\{\lambda\})|\phi(\{\lambda\})\rangle
  -\langle\phi(\{\lambda\})|\phi(\{\lambda\})\rangle   \nonumber\\
&&\mbox{}\,\,+\sum_{j=1}^{N}\lim_{v\rightarrow z_m^2}
   {v-z_m^2\over \lambda_j-z_m^2}
   \langle\phi(\{\lambda\})|
   {\cal C}(\lambda_1)\ldots {\cal C}(\lambda_{j-1}){\cal C}(v)
   {\cal C}(\lambda_{j+1})\ldots {\cal C}(\lambda_{N})|0\rangle \nonumber\\
&&=\mbox{det} {\tilde G}(\{\lambda\}) -
 \mbox{det}\left({\tilde G}(\{\lambda\}) -{\tilde Q}_m(\{\lambda\})
    \right) \nonumber
\end{eqnarray}
with the elements of the rank-one matrix ${\tilde Q}_m(\{\lambda\})$ being
\begin{eqnarray*}
\left({\tilde Q}_m(\{\lambda\})\right)_{ij} =\frac{z_m^2}{(\lambda_i-z_m^2)^2}.
\end{eqnarray*}
In the above derivation, we have used the following property of
determinants: If ${\cal A}$ is an arbitrary $n\times n$ matrix and
${\cal B}$ is a rank-one $n\times n$ matrix, then the determinant of
${\cal A}+{\cal B}$ is given by
$$
\mbox{det}({\cal A}+{\cal B})=\mbox{det}{\cal A}
 +\sum_{i=1}^n\mbox{det}{\cal A}^{(i)},
$$
where the elements of the matrix ${\cal A}^{(i)}$ are defined as
\begin{eqnarray*}
&&{\cal A}^{(i)}_{\alpha\beta}={\cal A}_{\alpha\beta} \quad\quad
 \mbox{ for } \beta\ne i,\\
&&{\cal A}^{(i)}_{\alpha i}={\cal B}_{\alpha i}.
\end{eqnarray*}

\begin{itemize}
\item Here we calculate the diagonal one-point function for the boson number operator $N_b$
\begin{eqnarray*}
F(\{\lambda\},N_b,\{\lambda\})=
\langle\phi(\{\lambda\})|{N}_{b}|\phi(\{\lambda\})\rangle.
\end{eqnarray*}
\end{itemize}
Similar to the above case,  using the commutation relation (\ref{eq:comm-NaC}) we obtain the one-point function for $N_b$
as 
\begin{eqnarray}
F(\{\lambda\},N_b,\{\lambda\})
&=&\sum_{j=1}^{N}{2z_b\over \lambda_j}
   \langle\phi(\{\lambda\})|b^\dag\prod_{k\ne j}^{N}
    {\cal C}(\lambda_k)|0\rangle \nonumber\\
&=&\langle\phi(\{\lambda\})|\phi(\{\lambda\})\rangle
  -\langle\phi(\{\lambda\})|\phi(\{\lambda\})\rangle   \nonumber\\
&&\mbox{}\,\,+\sum_{j=1}^{N}\lim_{v\rightarrow 0}
  {v\over\lambda_j}
   \langle\phi(\{\lambda\}_{N})|{\cal C}(v)\prod_{k\ne j}^{N}
    {\cal C}(\lambda_k)|0\rangle \nonumber\\
&=&\mbox{det}{\tilde G}(\{\lambda\})
-\mbox{det}\left({\tilde G}(\{\lambda\})
   -{\tilde Q}_b(\{\lambda\}) \right)
  \label{eq:cf-Nb}
\end{eqnarray}
with the elements of the rank-one matrix ${\tilde Q}_b(\{\lambda\})$ reading
\begin{eqnarray*}
 \left({\tilde Q}_b(\{\lambda\})\right)_{ij}={4z_b^2\over \lambda_i^2}.
\end{eqnarray*}
Through this last example we can compute the boson expectation defined by
$$
\langle
N_b\rangle={\langle\phi(\{\lambda\})|N_b|\phi(\{\lambda\})\rangle\over
  \langle\phi(\{\lambda\})|\phi(\{\lambda\})\rangle}.
$$
Substituting
(\ref{eq:sclr-pr-eq}) and (\ref{eq:cf-Nb}) into the above
definition, we obtain
\begin{eqnarray}
\langle N_b\rangle=1-{\mbox{det}\left({\tilde
G}(\{\lambda\})-{\tilde Q}_b(\{\lambda\})\right)\over
\mbox{det}\left({\tilde
G}(\{\lambda\})\right)}\label{eq:expect-Nb}
\end{eqnarray}
and in turn the boson fraction expectation value $\langle N_b\rangle/N$ which has previously been used to characterise BEC-BCS crossover properties \cite{abcg05}. 

\begin{figure}[t!]
\begin{center}
\includegraphics[height=8cm]{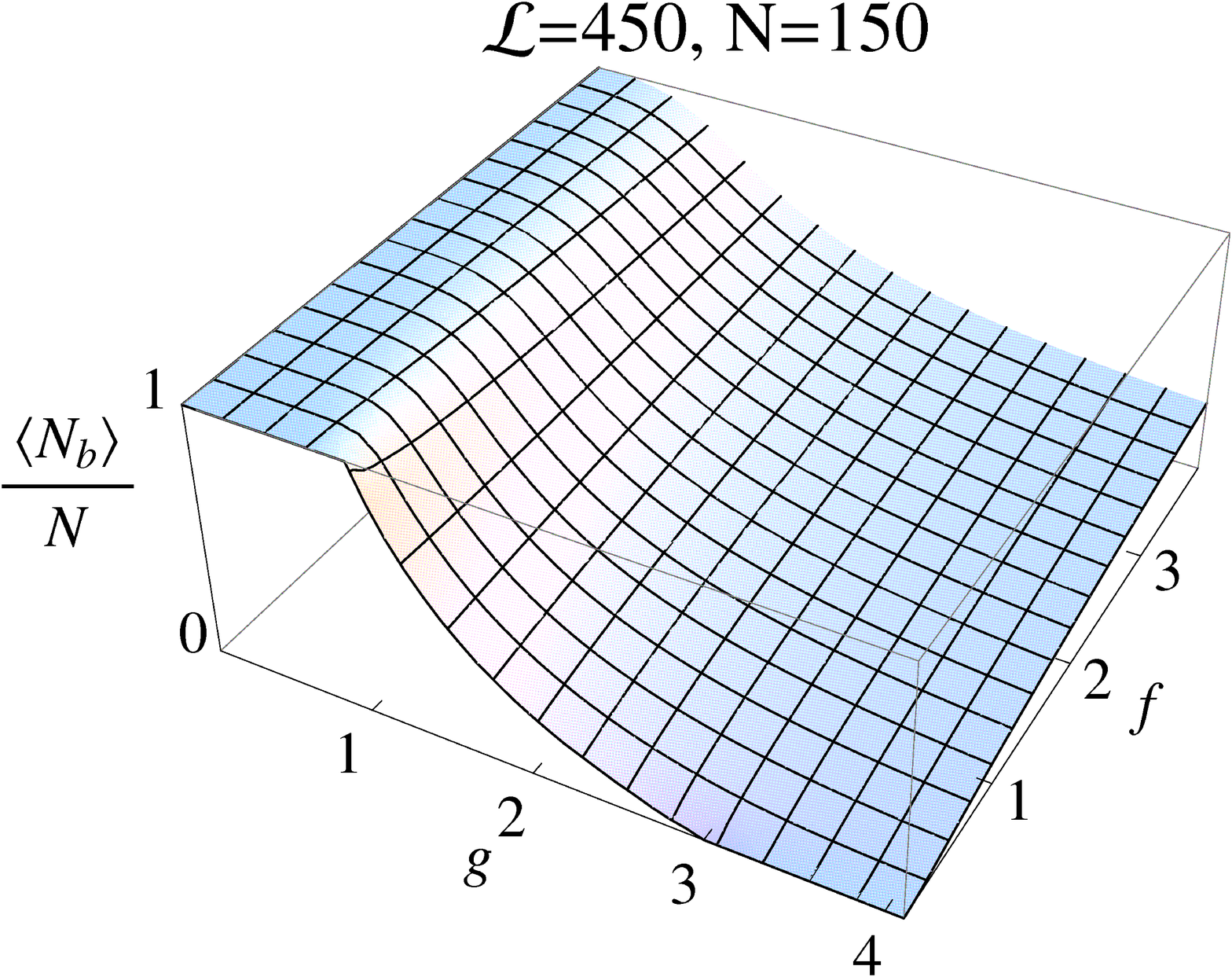}
\end{center}
\caption{The boson fraction expectation value
$\langle N_b \rangle/N$ as a function of the coupling parameters
$f$, $g$, as given by (\ref{eq:expect-Nb}). The results shown are
for a system of ${\cal L}=450$ momentum pair states and ${N}=150$
pairs, giving the filling fraction as $x=1/3$. The boson fraction
expectation value shows smooth variation between the  BCS ($\langle
N_b \rangle /N=0$) and BEC ($\langle N_b \rangle /N=1$) extremes. }
\label{fg:cf-Nb}
\end{figure}

Since the ground-state roots of the Bethe ansatz equations (\ref{BAE})
is the unique solution set which is real and negative, this makes for an
efficient study of the ground-state features in finite
systems. This is because the numerical solution of (\ref{BAE})
for negative real roots is very reliable, due to the uniqueness of
a solution with this property. As an example, we take $2{\cal L}=900$ momenta  which
arise in pairs $\bk$ and $-\bk$. The distribution of the momenta is
chosen as $|\bk|=\sqrt{2n},\,n=1,...,450$, giving the cut-off as
$\omega=30$. Taking the total particle number as
${\mathcal{N}}=300$, corresponding to the filling fraction $x=1/3$,
we numerically solve for the ground-state roots of the BAEs
(\ref{BAE}) to calculate (\ref{eq:expect-Nb}). In this sector the
Hilbert space has dimension $1.96\times 10^{123}$. The results shown
in Fig.\ref{fg:cf-Nb} suggest smooth variation of the boson fraction
expectation value for $f,\,g>0$, consistent with the absence of a phase transition.


%
%


\subsection{Two-point correlation functions}

\begin{itemize}
\item We first determine the off-diagonal two-point correlation function for $s_m^\dag b$
$$
F(\{\lambda\},s_m^\dag b,\{\lambda\})
 =\langle\phi(\{\lambda\})| s_m^\dag b
|\phi(\{\lambda\})\rangle.
$$
\end{itemize}
The commutation relation between operators $b $ and ${\cal
C}(\lambda)$, viz.
$$
[b ,{\cal C}(\lambda)]={2z_b\over \lambda}
$$
allows us to commute the bosonic operator with all ${\cal
C}(\lambda_j)$. Considering that $b|0\rangle=0$ we obtain
\begin{eqnarray}
F(\{\lambda\},s_m^\dag b,\{\lambda\})
&=&
   \sum_{j=1}^{N}{2z_b\over \lambda_j}
   \langle\phi(\{\lambda\})|s_m^\dag
   \prod_{k\ne j}^{N}{\cal C}(\lambda_k)|0\rangle
   \nonumber\\
&=&\langle\phi(\{\lambda\})|\phi(\{\lambda\})\rangle-
   \langle\phi(\{\lambda\})|\phi(\{\lambda\})\rangle \nonumber\\
&&\mbox{~~~}
   + \lim_{v\rightarrow z_m^2}{v-z_m^2\over z_m}
  \sum_{j=1}^{N}{2z_b\over \lambda_j}
   \langle\phi(\{\lambda\})|{\cal C}(v)
   \prod_{k\ne j}^{N}{\cal C}(\lambda_k)|0\rangle
   \nonumber\\
&=&\mbox{det}{\tilde G}(\{\lambda\})
-\mbox{det}\left({\tilde G}(\{\lambda\}) -{\tilde M}_m(\{\lambda\}) \right)
    \label{eq:cf-bs}
\end{eqnarray}
with the elements of the rank-one matrix ${\tilde M}_m$ as
\begin{eqnarray*}
 \left({\tilde M}_m(\{\lambda\}) \right)_{ij}=
 {2z_bz_m(\lambda_j-z_m^2)\over \lambda_j(\lambda_i-z_m^2)^2}.
\end{eqnarray*}

\begin{figure}[!h]
\begin{center}
\vspace{1cm}
$
\begin{array}{cc}
\epsfxsize=7cm \epsffile{ 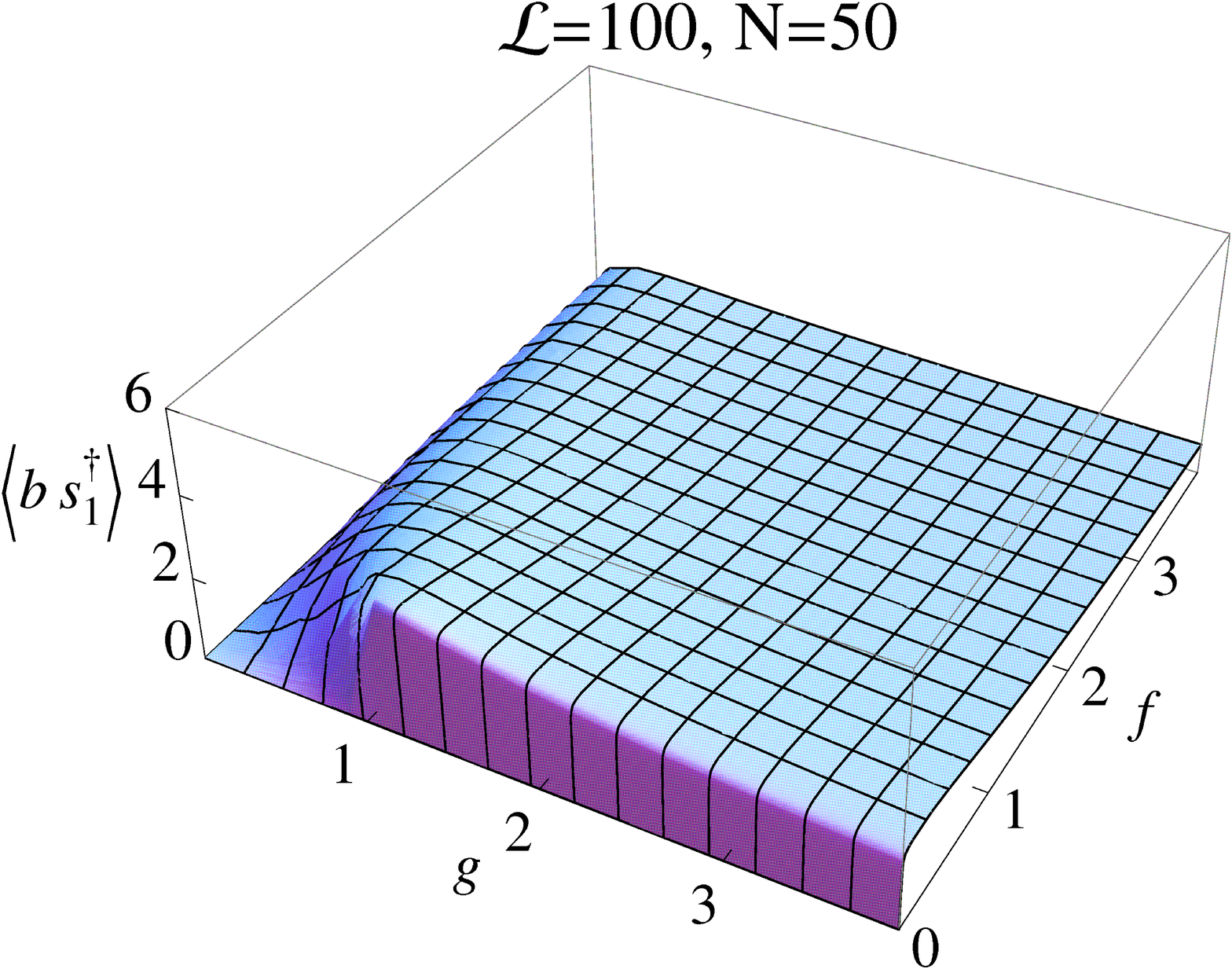 } & \epsfxsize=7cm
\epsffile{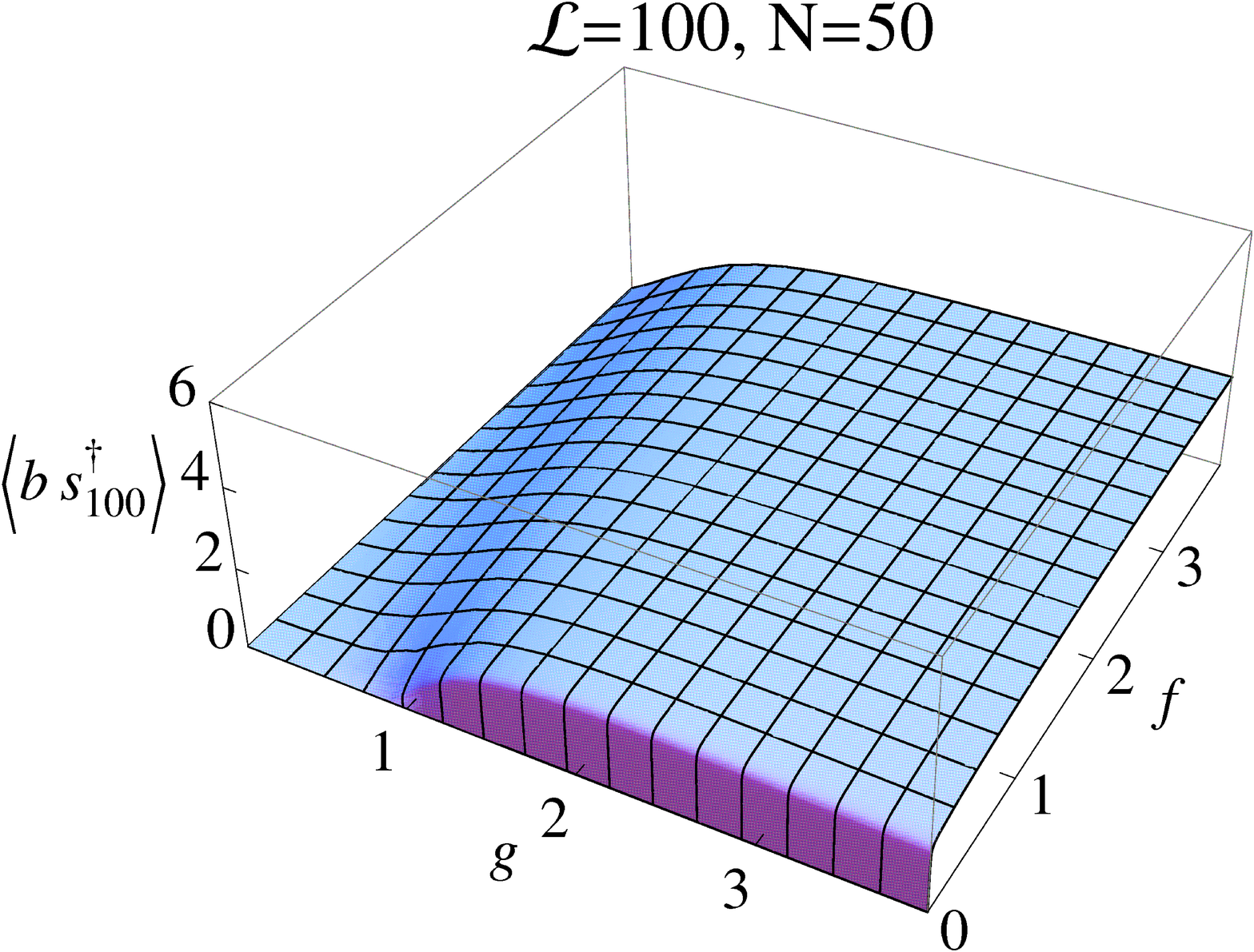} \\[0.4cm] \mbox{\bf (a)} & \mbox{\bf
(b)} \\[0.4cm] \epsfxsize=7cm \epsffile{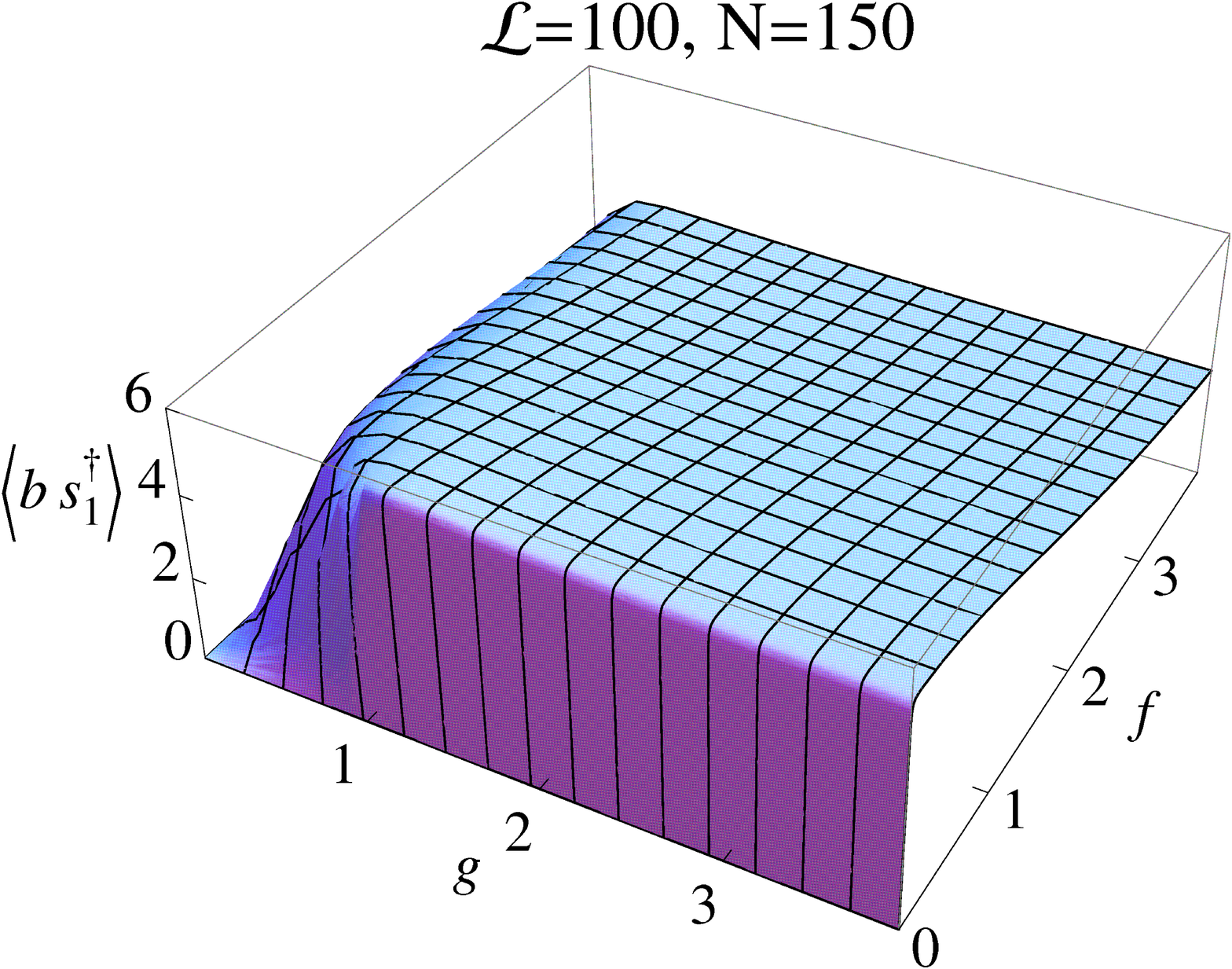} &
\epsfxsize=7cm \epsffile{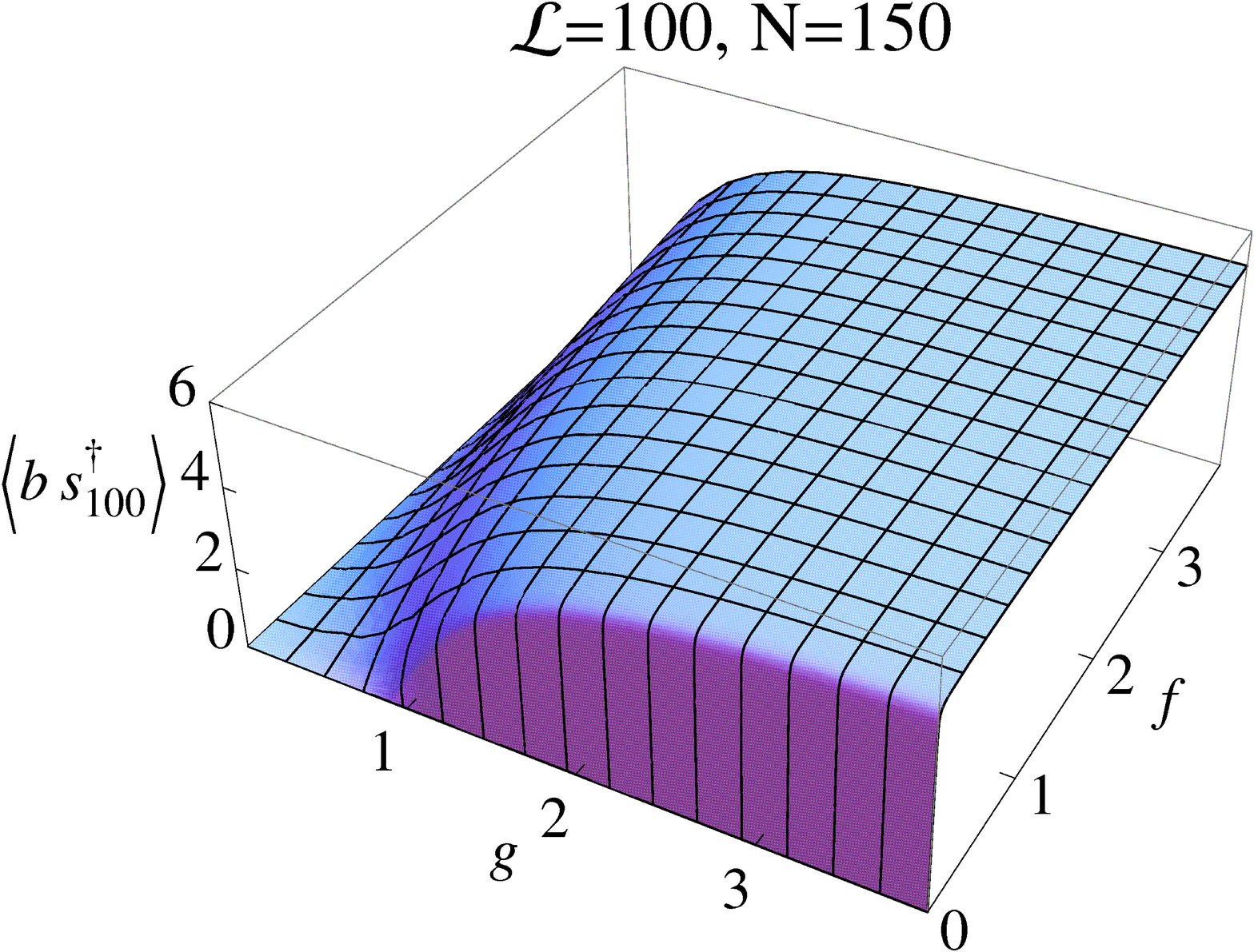} \\[0.4cm] \mbox{\bf (c)}
& \mbox{\bf (d)}
\end{array}
$
\end{center}
\vspace{1cm}
\caption{Ground-state two-point correlation function $\langle
bs_n^\dagger\rangle $ (\ref{eq:bsndagger}), as a function of the coupling parameters $f$
and $g$.  These two pinjt functions represent the boson-Cooper pair quantum
fluctuations in finite systems where $\langle
b\rangle =\langle
s_n^\dagger\rangle =0$, Here ${\cal L}=100$ and the distribution of the momenta is $|{\mathbf k}|=\sqrt{2n}$. The insets correspond to the cases
 (a) $N=50$, $n=1$; (b) $N=50$, $n=100$; (c) $N=150$;
$n=1$; (d) $N=150$, $n=100$. For intermediate values $1<n<100$ we
have found that $\langle bs_n^\dagger\rangle $ has the same generic
profile.  It is
apparent that increasing $N$ is associated with increasing
fluctuations.}\label{fg:cf-bs}
\end{figure}

\begin{figure}[!h]
\begin{center}
$
\begin{array}{cc}
\epsfxsize=7.5cm \epsffile{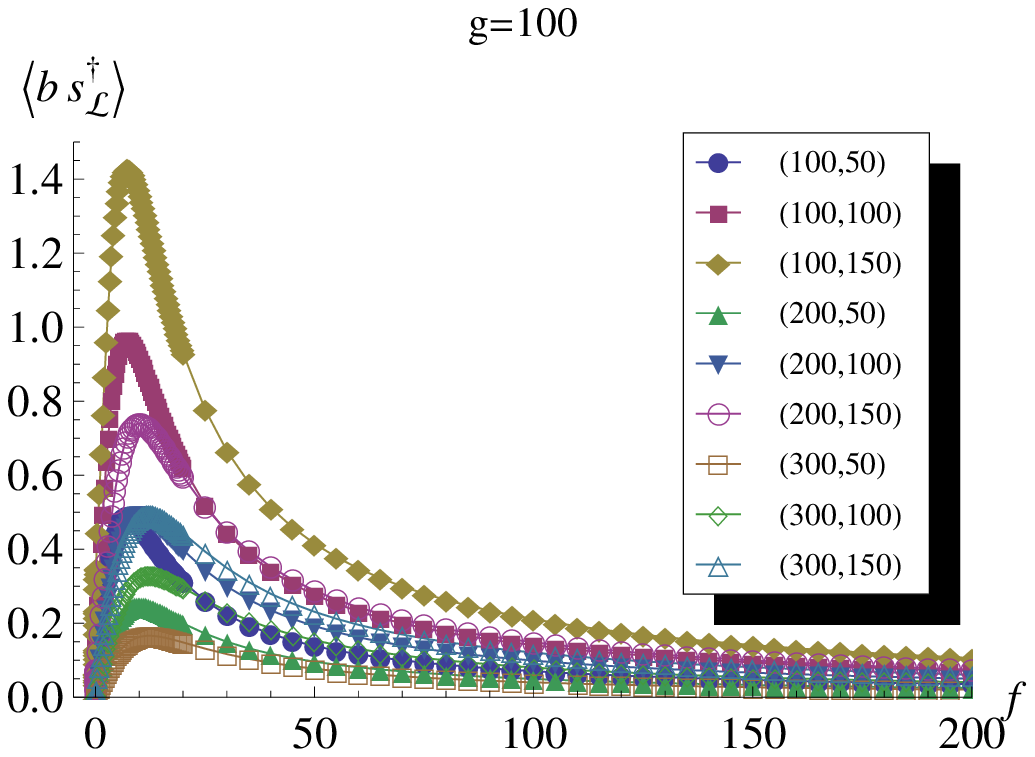}&
\epsfxsize=6cm\epsffile{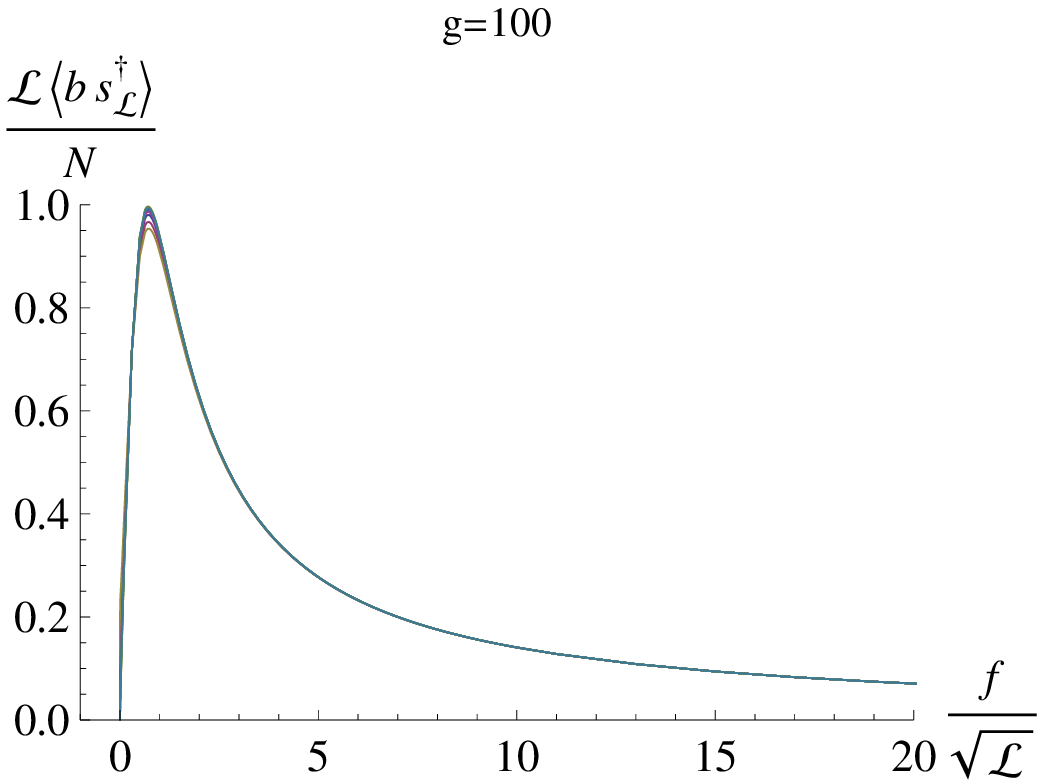} \\[0.4cm]
\mbox{\bf (a)} & \mbox{\bf(b)} \\[0.4cm]
\epsfxsize=6cm \epsffile{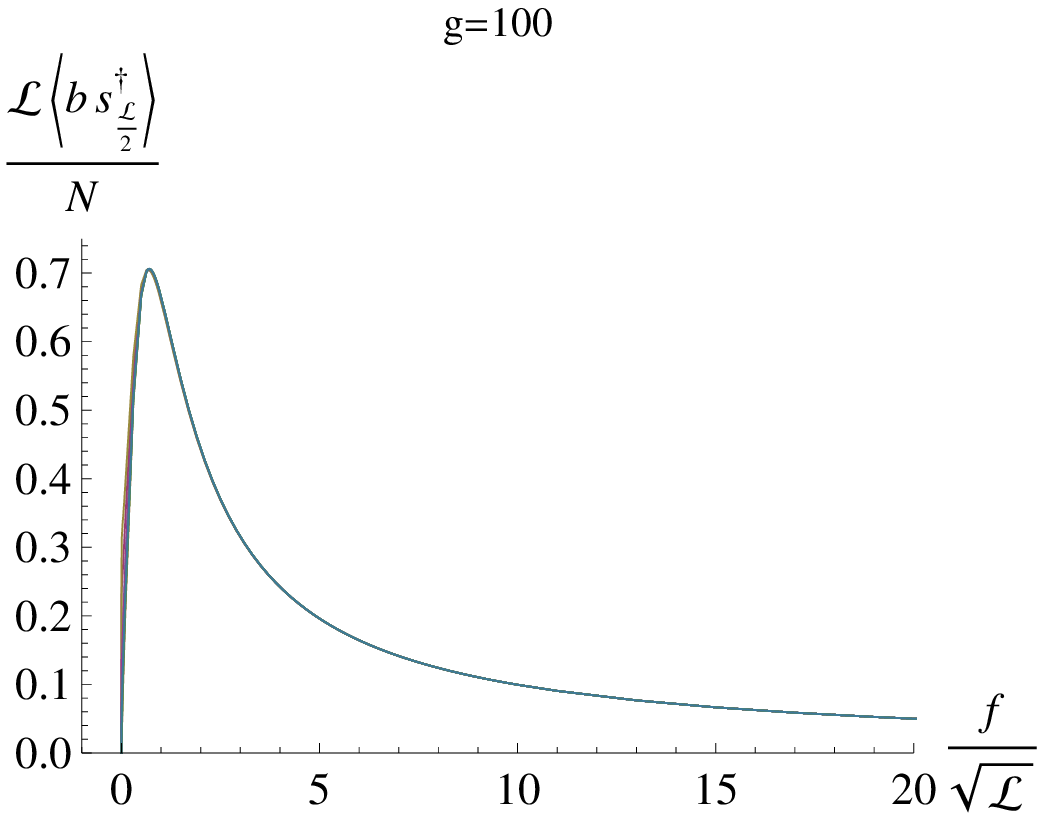} &
\epsfxsize=6cm \epsffile{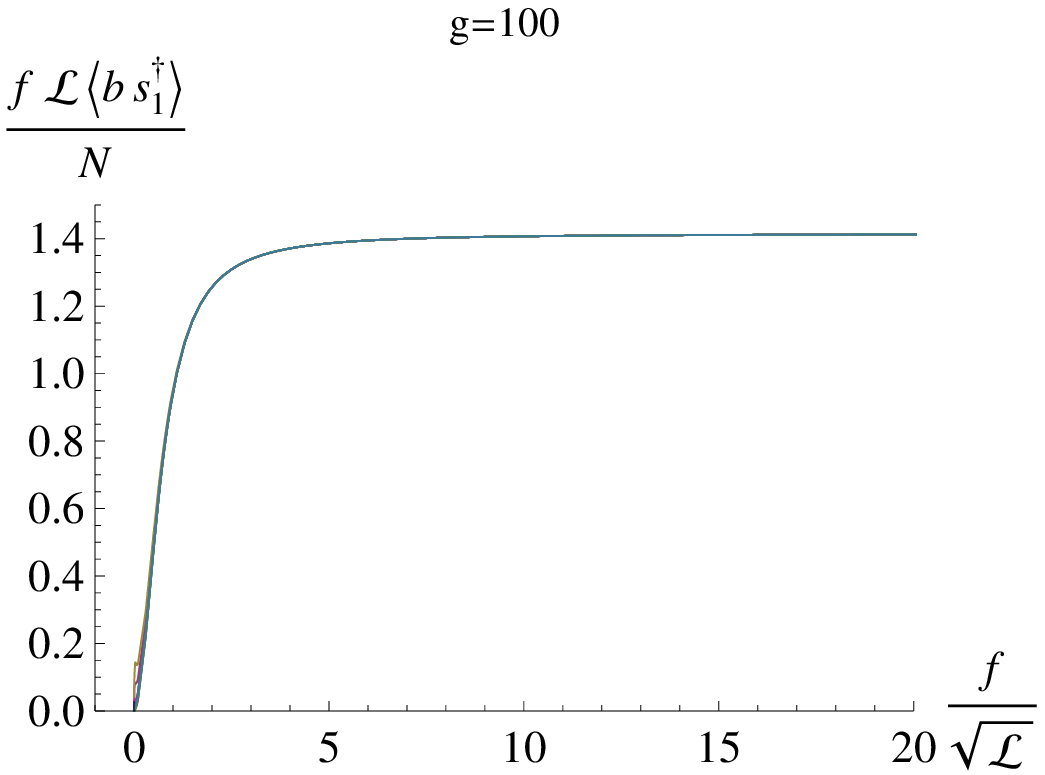} \\[0.4cm]
\mbox{\bf (c)} & \mbox{\bf (d)}
\end{array}
$

\end{center}
\caption{Relationships between the boson-Cooper pair quantum
fluctuations $\langle bs_n^\dag\rangle$ and the rescaled coupling
parameter $f/\sqrt{{\cal L}}$ with $g=100$, for different pairs of
$({\cal L},N)$. The momentum distribution has been rescaled as $|{\mathbf k}|=\sqrt{2n/\cal L}$. Inset (a) shows the
fluctuations for the case $n=\cal L$ as functions of $f$.
Here there are nine distinguishable curves corresponding to the choices of 
$({\cal L},N)$ where ${\cal L}=100,200,330$ and $N=50,100,150$. The remaining insets for (b) $n=\cal L$, (c) $n={\cal L}/2$, (d) $n=1$ illustrate that, with the
inclusion of scaling factors, the pair
quantum fluctuations can be expressed by functions of
the variable $f/\sqrt{\cal L}$. These cases also display data for the nine choices of $({\cal L},N)$, which is seen to fall on a single curve.  The scaled fluctuations go to zero as $f/\sqrt{\cal L}\rightarrow 0$. This indicates that the fluctuations vanish in the thermodynamic limit $N,\,{\cal L}\rightarrow \infty$, with $x=N/{\cal L}$ finite.
\vspace{2cm}}
\label{fg:scale}
\end{figure}

The canonical two-point function with the definition
$$
\langle bs^\dag_n\rangle=
 {\langle\phi(\{\lambda\})| s_n^\dag b|\phi(\{\lambda\})\rangle \over
 \langle\phi(\{\lambda\})|\phi(\{\lambda\}) \rangle}.
$$
is expressible, using (\ref{eq:cf-bs}), as 
\begin{eqnarray}
\langle bs^\dag_n\rangle=1-{\mbox{det}\left({\tilde
G}(\{\lambda\})-{\tilde M}_n(\{\lambda\})\right)\over
\mbox{det}\left({\tilde G}(\{\lambda\})\right)}.
\label{eq:bsndagger}
\end{eqnarray}
Fig. \ref{fg:cf-bs} illustrates the behaviour of the ground-state
two-point function $\langle b s_n^\dag\rangle$ as a function of the
coupling parameters $f,g$. In all instances there is a rapid decrease in the
fluctuations as $f\rightarrow 0$, but they appear smooth nonetheless for $f,g>0$.
Fig. \ref{fg:scale} shows the scaling behaviour of these two-point functions. These results indicate that, for a fixed value of $g$, we may write 
$$\langle b s_n^\dag\rangle = \phi_n(x,f)\theta_n(f/{\cal L})$$
where $\phi_n(x,f)$ is finite and $\theta_n(f/{\cal L})$ has the property that $\theta_n(0)=0$. For the thermodynamic limit $N,\,{\cal L}\rightarrow \infty$ with $x=N/{\cal L}$ we conclude that $\langle b s_n^\dag\rangle \rightarrow 0$, which is one of the main assumptions underlying the mean-field treatment discussed in the Appendix. 
    
For the remainder of this subsection we calculate three more cases of two-point correlation functions. Although we will not numerically evaluate these examples, the formulae are included for completeness. 
\begin{itemize}
\item Here we calculate the two-point correlation function for $s^\dag_m s_n$
$$
F(\{\lambda\},s_m^\dag s_n ,\{\lambda\})
 =\langle\phi(\{\lambda\})| s_m^\dag s_n 
|\phi(\{\lambda\})\rangle
$$
\end{itemize}

Operating $s_n$ on the state $|\phi(\{\lambda\})\rangle$, we have
$$
s_n|\phi(\{\lambda\}_N)\rangle=
 s_n\prod_{\beta=1}^N
 {\cal C}(\lambda_\beta)|0\rangle
=s_n\prod_{\beta=1}^N\left(\tilde {\cal C}_n(\lambda_\beta)+A_n^\beta
s^\dag_n\right)|0\rangle,
$$
where
$$ \tilde {\cal C}_n(\lambda_\beta)={2z_bb^\dagger \over \lambda_\beta}+
\sum_{l\ne n}^N
 {z_l s^\dag_l\over \lambda_\beta-z_l^2}, \quad\quad
 A_n^\beta={z_n\over \lambda_\beta-z_n^2}. $$
Bearing in mind that $(s^\dag_n)^2=0,\, \, s_n|0\rangle=0$, we find
that the non-zero terms in the above relation combine to give
 \begin{eqnarray}
s_n|\phi(\{\lambda\})\rangle &=& \sum_{\beta=1}^N
A_n^\beta\prod_{\alpha\ne\beta}^N
\tilde {\cal C}_n(\lambda_\beta)|0\rangle\nonumber\\
&=&\sum_{\beta=1}^NA_n^\beta\prod_{\alpha\ne\beta }^N
\left({\cal C}(\lambda_\alpha)-A_n^\alpha s^\dag_n\right)|0\rangle\nonumber\\
&=&\sum_{\beta=1}^NA_n^\beta\prod_{\alpha\ne \beta}^N{\cal
C}(\lambda_\alpha)|0\rangle
   -2\sum_{\beta=1}^N\sum_{\alpha<\beta}^N
   A_n^\alpha A_n^\beta s^\dag_n\prod_{\gamma\ne \alpha,\beta}^N
   {\cal C}(\lambda_\gamma)|0\rangle.
   \label{eq:b+}
\end{eqnarray}
With the aid of (\ref{eq:b+}), the two-point function reduces to
\begin{eqnarray*}
F(\{\lambda\},s^\dag_ms_n,\{\lambda\})
&=&\sum_{\beta=1}^N A_n^\beta\langle \phi(\{\lambda\})
 |s^\dag_m \prod_{\alpha\ne \beta}^N
 {\cal C}(\lambda_\alpha)|0\rangle
   \nonumber\\ &&\mbox{}\quad
 -2\sum_{\beta=1}^N\sum_{\alpha<\beta}^NA_n^\alpha A_n^\beta
  \langle \phi(\{\lambda\})|s^\dag_m s^\dag_n
  \prod_{\gamma\ne \alpha,\beta}^N{\cal C}(\lambda_\gamma)|0\rangle. 
\end{eqnarray*}
Now the two-point function for $s^\dag_ms_n$ has been simplified to
sums of one-point functions of $s^\dag_m$ and two-point functions of
$s^\dag_m s^\dag_n$. 
For the first term, we have
$$
\sum_{\beta=1}^N A_n^\beta\langle \phi(\{\lambda\})
 |s^\dag_m \prod_{\alpha\ne \beta}^N
 {\cal C}(\lambda_\alpha)|0\rangle
=\sum_{\beta=1}^N \left[
A_n^\beta(\lambda_\beta-z_m^2)
\right]
\mbox{det} {\tilde T}_m^\beta(\{\lambda\}),
$$
where
\begin{eqnarray}
\left( {\tilde T}_m^\beta(\{\lambda\}) \right)_{ij} 
&=& \left({\tilde G}(\{\lambda\})\right)_{ij},\quad j\neq \beta,\nonumber\\
\left( {\tilde T}_m^\beta(\{\lambda\}) \right)_{i\beta} &=& 
\frac{z_m}{(\lambda_i - z_m^2)^2}.
\nonumber
\end{eqnarray}
For the second term, we have
\begin{eqnarray}
&&2\sum_{\beta=1}^N\sum_{\alpha<\beta}^NA_n^\alpha A_n^\beta
  \langle  \phi\{\lambda\})|s^\dag_m s^\dag_n
  \prod_{\gamma\ne \alpha,\beta}^N
  {\cal C}(\lambda_\gamma)|0\rangle
 \nonumber\\ &&
 =\lim_{u\rightarrow z_m^2}\lim_{v\rightarrow z_n^2}{(u-z_m^2)(v-z_n^2)\over z_mz_n}
 \sum_{\beta=1}^N\sum_{\alpha<\beta}^N2A_n^\alpha A_n^\beta
 \langle \phi\{\lambda\})|{\cal C}(u){\cal C}(v)
 \prod_{\gamma\ne \alpha,\beta}^N{\cal C}(\lambda_\gamma)|0\rangle
  \nonumber\\
&&=
  \sum_{\beta=1}^N \left( \left[ A_n^\beta(\lambda_\beta-z_m^2)
\right]
\sum_{\alpha< \beta}^N   K_{mn}^{\alpha\beta}
   \mbox{det}({\tilde T}_{mn}^{\alpha\beta}(\{\lambda\})) \right) \nonumber
\end{eqnarray}
with
$$
({\tilde T}_{mn}^{\alpha\beta}(\{\lambda\}))_{ij}=\left({\tilde G}(\{\lambda\})\right)_{ij},\qquad j\ne \alpha,\beta,
$$
$$
({\tilde T}_{mn}^{\alpha\beta}(\{\lambda\}))_{i\alpha}= \frac{z_n}{(\lambda_i-z_n^2)^2},\quad \quad
({\tilde T}_{mn}^{\alpha\beta}(\{\lambda\}))_{i\beta}= \frac{z_m}{(\lambda_i-z_m^2)^2},
$$%
$$
K_{mn}^{\alpha\beta}=
 {2z_n(\lambda_\alpha-z_m^2)(\lambda_\beta-z_n^2)\over
 (\lambda_\alpha-\lambda_\beta)(z^2_m-z^2_n)}.
$$
We can therefore express this two-point function as
\begin{eqnarray}
&& F(\{\lambda\},s^\dag_ms_n,\{\lambda\})\nonumber\\
&& \ \ = 
\sum_{\beta=1}^N 
\left[
A_n^\beta(\lambda_\beta - z_m^2)
\right]
\left( 
\mbox{det} {\tilde T}_m^\beta(\{\lambda\})
-
\sum_{\alpha<\beta}
K_{mn}^{\alpha\beta}
\mbox{det}({\tilde T}_{mn}^{\alpha\beta}(\{\lambda\}))
 \right).
\nonumber
\end{eqnarray}
Writing the columns of the matrices ${\tilde T}_m^\beta$ and ${\tilde T}_{mn}^{\alpha\beta}$ in
vector notation, we have
\begin{eqnarray}
\mbox{det}({\tilde T}_{m}^{\beta})&=&
 \mbox{det}\left(\vec { G}_1,\ldots,
  \vec { G}_{\beta-1}, \vec { W}_m,
  \vec { G}_{\beta+1}, \ldots,
  \vec { G}_N\right),\nonumber\\
\mbox{det}({\tilde T}_{mn}^{\alpha\beta})&=&
 \mbox{det}\left(\vec { G}_1,\ldots,
  \vec { G}_{\alpha-1}, \vec { W}_n,
  \vec { G}_{\alpha+1} \ldots\,
  \vec { G}_{\beta-1}, \vec { W}_m,
  \vec { G}_{\beta+1}, \ldots,
  \vec { G}_N\right),\nonumber
\end{eqnarray}
where $\vec { G}_j$, $j=1,\ldots,N$ 
denotes the $N$-dimensional vector with entries $(\vec { G}_j)_i={\tilde G}_{ij}$
and $\vec { W}_x$ $(x=m,n)$ denotes the $N$-dimensional vector with entries 
$$
(\vec {W}_x)_i=\frac{z_x}{(\lambda_i-z_x^2)^2}.
$$
Note that we have refrained from detailing the explicit dependence on
$\{\lambda\}$ in the above expression, and hope it is still clear to the
reader.
We focus our attention on simplifying the expression
$$
\mbox{det} {\tilde T}_m^\beta
-
\sum_{\alpha<\beta}
K_{mn}^{\alpha\beta}
\mbox{det}{\tilde T}_{mn}^{\alpha\beta}
$$
for each permissible $\beta.$ Using properties of determinants, it is
possible to establish that
$$
\mbox{det} {\tilde T}_m^\beta
-
\sum_{\alpha<\beta}
K_{mn}^{\alpha\beta}
\mbox{det}{\tilde T}_{mn}^{\alpha\beta}
=
\mbox{det}{\tilde X}_{mn}^{\beta}
$$
where
\begin{eqnarray}
\left( {\tilde X}_{mn}^{\beta} \right)_{ij} & = & {\tilde G}_{ij} -
K_{mn}^{j\beta}\frac{z_n}{(\lambda_i-z_n^2)^2},\ \ j<\beta,
\nonumber\\
\left( {\tilde X}_{mn}^{\beta} \right)_{i\beta} & = & \frac{z_m}{(\lambda_i-z_m^2)^2},
\nonumber\\
\left( {\tilde X}_{mn}^{\beta} \right)_{ij} & = & {\tilde G}_{ij}, \ \ j>\beta.
\nonumber
\end{eqnarray}
Therefore, the two-point function simplifies to
\begin{eqnarray}
F(\{\lambda\},s^\dag_ms_n,\{\lambda\})
&=& 
\sum_{\beta=1}^N 
\left[
\frac{z_n(\lambda_\beta - z_m^2)}{\lambda_\beta-z_n^2}
\right]
\mbox{det}{\tilde X}_{mn}^{\beta}(\{\lambda\})
\nonumber
\end{eqnarray}
We remark that the above procedure for reducing the double sum of determinants to a single sum leads to a more compact expression compared to the analogous result in \cite{dilsz10}.

\begin{itemize}
\item Two-point function of ${N}_{m}{N}_{n}$
\end{itemize}

Now we consider the two-point function of ${N}_{m}{N}_{n}$
$$
F(\{\lambda\},{N}_{m}{N}_{n},\{\lambda\}) =\langle\phi(\{\lambda\})|
{N}_{m}{N}_{n}
 \prod_{i=1}^N {\cal C}(\lambda_i)|0\rangle.
$$
Commuting the number operators by using the commutation relation
(\ref{eq:comm-NC}), we derive the correlation function as follows:
\begin{eqnarray*}
F(\{\lambda\},{N}_{m}{N}_{n},\{\lambda\}) 
 & =&\langle\phi(\{\lambda\})|{N}_{m}{N}_{n}
 \prod_{i=1}^N {\cal C}(\lambda_i)|0\rangle \nonumber\\
&=&\sum_{\beta=1}^N{z_m\over \lambda_\beta-z_m^2}
 \langle\phi(\{\lambda\})| {N}_{n} s_m^\dag
 \prod_{\alpha\ne\beta}^N {\cal C}(\lambda_\alpha)|0\rangle \nonumber\\
&=&\sum_{\beta=1}^N{z_m\over \lambda_\beta-z_m^2}
   \sum_{\alpha\ne\beta}^N{z_n\over \lambda_\alpha-z_n^2}
 \langle\phi(\{\lambda\})|s_m^\dag s_n^\dag
 \prod_{\gamma\ne\alpha,\beta}^N {\cal C}(\lambda_\gamma)|0\rangle \nonumber\\
&=&
  \sum_{\beta=1}^N\sum_{\alpha\ne \beta}^N
   J_{mn}^{\alpha\beta}\mbox{det}({\tilde T}_{mn}^{\alpha\beta})\nonumber\\
&=&
\sum_{\beta=1}^N\sum_{\alpha<\beta}^N(J_{mn}^{\alpha\beta}-J_{mn}^{\beta\alpha})
\mbox{det}({\tilde T}_{mn}^{\alpha\beta})\nonumber\\
&=&
\sum_{\beta=2}^N\left[ \mbox{det}({\tilde T}_m^\beta) -  \mbox{det}({\tilde T}_m^\beta)\right] + 
\sum_{\beta=1}^N\sum_{\alpha<\beta}^N(J_{mn}^{\alpha\beta}-J_{mn}^{\beta\alpha})
\mbox{det}({\tilde T}_{mn}^{\alpha\beta}),\nonumber\\
\end{eqnarray*}
where
$$
J_{mn}^{\alpha\beta}=
 {z_mz_n(\lambda_\alpha-z_m^2)(\lambda_\beta-z_n^2)\over
 (z^2_m-z^2_n)(\lambda_\alpha-\lambda_\beta)}.
$$
At this stage it is worth pointing out the obvious fact that in the second term in the last
line of calculation above, the summation never sees the $\beta =1$ term, so we can proceed by writing
\begin{eqnarray*}
F(\{\lambda\},{N}_{m}{N}_{n},\{\lambda\}) 
 & =&
\sum_{\beta=2}^N\left[ \mbox{det}({\tilde T}_m^\beta) +\sum_{\alpha<\beta}^N(J_{mn}^{\alpha\beta}-J_{mn}^{\beta\alpha})
\mbox{det}({\tilde T}_{mn}^{\alpha\beta}) \right] - \sum_{\beta=2}^N \mbox{det}({\tilde T}_m^\beta).
\end{eqnarray*}
For each $\beta$, using similar techniques as before, the expression in square brackets above can be simplified to a single
determinant, namely
$$
\mbox{det}({\tilde T}_m^\beta) +\sum_{\alpha<\beta}^N(J_{mn}^{\alpha\beta}-J_{mn}^{\beta\alpha})
\mbox{det}({\tilde T}_{mn}^{\alpha\beta}) = 
\mbox{det}({\tilde Y}_{mn}^\beta),
$$
where the matrix elements are
\begin{eqnarray*}
\left({\tilde Y}_{mn}^\beta\right)_{ij} & = & {\tilde G}_{ij} + (J_{mn}^{j\beta}-J_{mn}^{\beta j})\frac{z_n}{(\lambda_i-z_n^2)^2},\ \ j<\beta,\\
\left({\tilde Y}_{mn}^\beta\right)_{i\beta} & = & \frac{z_m}{(\lambda_i-z_m^2)^2},\\
\left({\tilde Y}_{mn}^\beta\right)_{ij} & = & {\tilde G}_{ij},\ \ j>\beta.
\end{eqnarray*}
Once again using familiar properties of the determinant, we may also simplify
\begin{eqnarray*}
\sum_{\beta=2}^N \mbox{det}({\tilde T}_m^\beta) = \mbox{det}(A_m),
\end{eqnarray*}
where the matrix $A_m$ has elements given by
\begin{eqnarray*}
(A_m)_{i1} & = & {\tilde G}_{i1},\\
(A_m)_{ij} & = & {\tilde G}_{ij}-{\tilde G}_{i j+1},\ \ 1<j<N,\\
(A_m)_{iN} & = & \frac{z_m}{(\lambda_i-z_m^2)^2}.
\end{eqnarray*}
In the above we have supressed the explicit dependency on $\{\lambda\}$ in each of the
expressions, as it should be clear.
Therefore the two-point function can be expressed as a sum of $N$ determinants
$$
F(\{\lambda\},{N}_{m}{N}_{n},\{\lambda\}) 
  = \sum_{\beta=2}^N\mbox{det}({\tilde Y}_{mn}^\beta(\{\lambda\})) - \mbox{det}(A_m(\{\lambda\})).
$$


\begin{itemize}
\item Two-point function of $N_b {N}_{m}$
\end{itemize}

To compute the two-point function
$$
F(\{\lambda\},N_b {N}_{m},\{\lambda\}) 
 =\langle\phi(\{\lambda\})|N_b {N}_{m}
 \prod_{i=1}^N {\cal C}(\lambda_i)|0\rangle,
$$
we need to use  commutation relations both (\ref{eq:comm-NaC}) and
(\ref{eq:comm-NC}). The result is given by
\begin{eqnarray}
&&F(\{\lambda\},N_b {N}_{m},\{\lambda\}) 
 =\sum_{\beta=1}^N{z_m\over \lambda_\beta-z_m^2}
   \sum_{\alpha\ne\beta}^N{2z_b\over \lambda_\alpha}
 \langle\phi(\{\lambda\})|s_m^\dag b ^\dag
 \prod_{\gamma\ne\alpha,\beta}^N {\cal C}(\lambda_\gamma)|0\rangle \nonumber\\
&&=\lim_{u\rightarrow z_m^2}
 \sum_{\beta=1}^N{u-z_m^2\over \lambda_\beta-z_m^2}
 \lim_{v\rightarrow 0}\sum_{\alpha\ne\beta}^N{v\over \lambda_\alpha}
 \langle\phi(\{\lambda\})|{\cal C}(u){\cal C}(v)
 \prod_{\gamma\ne\alpha,\beta}^N {\cal C}(\lambda_\gamma)|0\rangle \nonumber\\
&&=
   \sum_{\beta=1}^N\sum_{\alpha\ne \beta}^N
   \tilde J_{mn}^{\alpha\beta}
   \mbox{det}({\tilde D}_{mn}^{\alpha\beta})\nonumber\\
&&=\sum_{\beta=1}^N\sum_{\alpha<\beta}(\tilde J_{mn}^{\alpha\beta} - \tilde
J_{mn}^{\beta\alpha}) \mbox{det}({\tilde D}_{mn}^{\alpha\beta})\nonumber
\end{eqnarray}
where
\begin{eqnarray*}
&&\tilde J_{mn}^{\alpha\beta}=
 {2z_b\lambda_\beta(\lambda_\alpha-z_m^2)\over
 z_m(\lambda_\alpha-\lambda_\beta)},
\nonumber\\
&& ({\tilde D}_{mn}^{\alpha\beta})_{ij}={\tilde G}_{ij}~~~
  (j\ne \alpha,\beta),~~~~~%
(\tilde D_{mn}^{\alpha\beta})_{i\alpha}= {2z_b\over \lambda_j^2}, \quad\quad 
(\tilde D_{mn}^{\alpha\beta})_{i\beta}=\frac{z_m}{(\lambda_i-z_m^2)^2}.
\end{eqnarray*}
Using similar techniques as before, we can reduce the above to a sum
of $N$ determinants. Doing this leads to
$$
F(\{\lambda\},{N}_b{N}_{m},\{\lambda\})\nonumber\\ =
\sum_{\beta=2}^N\mbox{det}({\tilde Z}_{mn}^\beta(\{\lambda\})) - \mbox{det}(A_m(\{\lambda\}))
$$
where the elements of ${\tilde Z}^{\beta}_{mn}$ are given by 
\begin{eqnarray*}
\left({\tilde Z}_{mn}^\beta\right)_{ij} & = & {\tilde G}_{ij} + 2({\tilde J}_{mn}^{j\beta}-
{\tilde J}_{mn}^{\beta j})\frac{z_b}{\lambda_i^2},\ \ j<\beta,\\
\left({\tilde Z}_{mn}^\beta\right)_{i\beta} & = & \frac{z_m}{(\lambda_i-z_m^2)^2},\\
\left({\tilde Z}_{mn}^\beta\right)_{ij} & = & {\tilde G}_{ij},\ \ j>\beta.
\end{eqnarray*}

\section{Conclusion}
We have introduced a model that couples a $p+ip$-wave pairing BCS Hamiltonian 
to a bosonic degree of freedom, and studied its properties regarding the 
BEC-BCS crossover. For a restriction on the coupling parameters, the model
was shown to be integrable and the exact solution was derived by the 
algebraic Bethe ansatz. We found that the ground-state roots of the Bethe ansatz equations  
have the property that they are the unique solution set such that all roots are real and negative.
Using this result we reasoned that the ground-state wavefunction is topologically trivial, so the 
BEC-BCS crossover is smooth. This conclusion was supported by a study of the boson fraction 
expectation value, which was computed exactly in the Bethe ansatz framework. We also formulated expressions for a range of two-point correlation functions and used one particular example to study the boson-Cooper pair fluctuations.
The range of correlation function expressions we have obtained provide ample scope for further studies along the lines of \cite{ao02}.

~~\\
~~\\
{\bf Acknowledgments}  C.D. and J.L. were funded through the
Royal Society Travel Grants Scheme. C.D. acknowledges support through EPSRC grant EP/G039526/1. P.S.I. was supported by an Early
Career Researcher Grant from The University of Queensland. J.L. and
S.-Y.Z. received funding by the Australian Research Council through
Discovery Project DP0663772. J.L. also acknowledges support through 
Discovery Project DP110101414.   

\section{Appendix - Mean-field theory}


In the mean-field approach products of operators $A$ and $B$ are approximated by
\begin{eqnarray}
AB \approx A\langle B \rangle + \langle A \rangle B - \langle A \rangle\langle B \rangle
\label{mf}
\end{eqnarray}
where the notation $\langle\cdot\rangle$ stands for the
expectation value.  This approximation assumes that quantum fluctuations may be neglected.
Formally, we define the fluctuations to be
$$
\chi(A,B)=\left|\langle AB\rangle -\langle A \rangle \langle B \rangle \right|
$$
such that within the mean-field approximation (\ref{mf}) we have
$$
\chi(A,B)=0.
$$
Applying
(\ref{mf}) to (\ref{de:Hamiltonian-z}) we obtain the mean-field Hamiltonian\footnote{We omit
the term $GH_0$ which becomes negligible in the thermodynamic limit which is discussed in
Subsection \ref{tl}.}
\begin{eqnarray}
H_{MF}&=&H_0+\left({K^2\over G}+\delta\right) N_b
 -{1\over 2G}\tilde\Delta^* (GQ+Kb)
 -{1\over 2G}\tilde\Delta (GQ^\dag+Kb^\dag)\nonumber\\&&\mbox{}
  +{1\over 4G}|\tilde\Delta|^2
 -\nu(N-\langle N\rangle),
     \label{de:Hamiltonian-MF}
\end{eqnarray}
where $\Delta=2G\langle Q \rangle+2K\langle b\rangle$ is referred to as the
gap. Since (\ref{de:Hamiltonian-MF}) does not commute with ${N}$ the Lagrange multiplier
$\nu$, the chemical potential, has been introduced in order to tune the expectation value
$\langle {N} \rangle$.

We take the following form for the mean-field variational ground state 
 $$|\Psi\rangle=|\alpha\rangle \otimes |\psi_1\rangle \otimes
 |\psi_2\rangle \otimes .... \otimes |\psi_{\cal L}\rangle, $$
 where $|\alpha\rangle$ is the coherent state such that
$b|\alpha\rangle=\alpha|\alpha\rangle$ with
$\alpha=|\alpha|e^{i\phi}\in {\mathbb C}$, and $|\psi_j\rangle$ are
local states related to the pairing operators
 \begin{eqnarray*}
 |\psi_j\rangle &=&(u_jI +v_j s^\dagger_j)|0_j\rangle
 \end{eqnarray*}
with $|0_j\rangle$ the vacuum state of the momentum pair space labelled by $j$.
Minimising the ground-state energy expectation value and imposing self-consistency of other operator expectation values leads to the following set of
equations determining the values for $\alpha$, the gap $\Delta$, and the chemical potential $\nu$:
\begin{eqnarray}
\alpha&=&\frac{K\Delta}{2(G\delta-G\nu+K^2)},
\label{mf3}\\
\frac{\delta-\nu}{G\delta-G\nu+K^2} &=& \sum_{j=1}^{\cal L}
\frac{z_j^2}{\sqrt{(z_j^2-\nu)^2+z_j^2|\Delta|^2}},
\label{mf1} 
\end{eqnarray}
\begin{eqnarray}
&&2{N}-{\cal L}-\frac{K\Delta^2}{2(G\delta-G\nu+K^2)^2}+
\frac{\delta-\nu}{G\delta-G\nu+K^2} \nonumber\\ 
&& \qquad\qquad\qquad\quad = \,\nu\sum_{j=1}^{\cal L}
\frac{1}{\sqrt{(z_j^2-\nu)^2+z_j^2|\Delta|^2}}.
\label{mf2}
\end{eqnarray}
The ground state energy assumes the form
\begin{eqnarray}
&&E = \nu|\alpha|^2+\frac{1}{2}\sum_{j=1}^{\cal L} z_j^2\left(1-
\frac{2z_j^2+|\Delta|^2-2\nu}{2\sqrt{(z_j^2-\nu)^2+z_j^2|\Delta|^2}}
\right) \label{mf4}
\end{eqnarray}
and we also find
\begin{eqnarray*}
|u_j|^2&=&\frac{1}{2}\left(1+\frac{z_j^2-\nu}{\sqrt{(z_j^2-\nu)^2+z_j^2|\Delta|^2}}  \right),  \\
 |v_j|^2&=&\frac{1}{2}\left(1-\frac{z_j^2-\nu}{\sqrt{(z_j^2-\nu)^2+z_j^2|\Delta|^2}}
 \right)=\langle {N}_{j} \rangle.
\end{eqnarray*}

Let $\nu^2=ab$ and $2\nu-|\Delta|^2=a+b$. It can be verified that taking the continuum limit for
(\ref{mf1})-(\ref{mf4}) with the substitution
(\ref{de:manifold}) reproduces  
equations (\ref{chempot})-(\ref{intnrg}).
Moreover when (\ref{de:manifold}) holds,  (\ref{mf3})-(\ref{mf2}) lead to the result
\begin{eqnarray*}
\frac{\langle N_b\rangle}{N} ={|\alpha|^2\over N}=
\frac{f^2|\Delta|^2}{16x\nu}=1-\frac{1}{2x}+\frac{1}{2x}
\int^\omega_0 d\vep \, \frac{\rho(\vep)(\vep+\sqrt{ab})}{R(\vep)}.
\end{eqnarray*}
This quantity is a smooth function for $f,g>0$, which can be shown
by using the fact that $a,b<0$. It displays discontinuous behaviour
(in the first derivative) only in the non-interacting limit
$f\rightarrow 0$ when $g=1$ or $g^{-1}=1-2x$, which is due to level
crossing. These correspond to the transition points of Table 2, and
are visible in Fig. 1. The smoothness of the boson fraction
expectation value for the interacting system in the thermodynamic
limit is consistent with the absence of a phase transition between
the BCS and BEC extremes.

\end{document}